\documentclass{JFM-FLM_Au}

\newcommand{\Wi}{\textit{Wi}}
\newcommand{\Bo}{\textit{Bo}}
\newcommand{\St}{\textit{St}}

\lefttitle{A. Woodbridge, C. P. Fonte and A. Juel}
\righttitle{Journal of Fluid Mechanics}

\title{Spreading dynamics of a drop of yield-stress fluid subject to vertical oscillations}

\author{Alice Woodbridge\aff{1},
  Cl\'{a}udio P. Fonte\aff{2}
 \and Anne Juel\aff{1}}

\affiliation{\aff{1}Department of Physics \& Astronomy, School of Natural Sciences, The University of Manchester, Oxford Road, Manchester M13 9PL, UK
\aff{2}Department of Chemical Engineering, School of Engineering, The University of Manchester, Manchester, M13 9PL, UK}

\corresau{Anne Juel, anne.juel@manchester.ac.uk}

\begin{document}
\maketitle

\begin{abstract}
The measurement of rheological properties from non-idealised flows is routinely used to characterise yield-stress materials. We extend the canonical slump test, which evaluates yield stress based on the spread of a large drop under gravity, by subjecting such a drop to vertical oscillation of its substrate. We test this flow configuration by direct comparison between experiments and numerical simulations of the time-dependent Cauchy equations coupled to the Saramito-Herschel-Bulkley (SHB) constitutive model. This provides a sensitive test of the SHB model near yield because the vibrated drop can releases stress exceeding the yield stress by adjusting its shape as the forcing acceleration is increased, and thus, remains close to the yield threshold for all forcing. We find that the SHB model systematically underpredicts the amplitude of the viscoelastic oscillatory response of the drop to forcing by a factor of five. However, spreading can be captured numerically by selecting a solvent viscosity that represents the appropriate post-yield rheology of the material at the expense of matching sub-yield properties. This highlights a fundamental challenge in simulating flows in which yielded and unyielded material coexist, such as those encountered in proto-rheological tests. We establish the vibrated drop as a powerful proto-rheological configuration for estimating the yield stress across drops of different rheology, size and shape, by establishing a proportional relationship between the threshold acceleration required to induce spreading relative to gravity and the yield number, the ratio of yield stress to gravitational stress acting at the centre of mass of the drop.  
\end{abstract}


\section{Introduction}\label{sec:Introduction}

Yield-stress fluids (YSFs) are among the most common and versatile materials encountered in everyday life. They are characterised by a yield stress that must be exceeded to initiate flow, and their dual solid–fluid nature makes them uniquely suited to applications that require structural integrity at rest and flow under stress \citep{Balmforth2014}, with examples ranging from food and personal care products to construction materials. Although they span a wide range of microstructures, their macroscopic flow behaviour is commonly shear-thinning with viscoelastic and sometimes thixotropic properties \citep{BONN2009, Dinkgreve2018}. The rheological characterisation of laboratory YSFs is usually performed under well-controlled, unidirectional shear flow using a rotational rheometer, by measuring the material response to an imposed stress, strain or strain rate. However, not all YSFs are amenable to characterisation in a rheometer, and in an industrial context, their material parameters may need to be estimated in-situ \citep{Tasaka}. 

A widely used alternative is proto-rheology where observation of less idealised flow configurations is used to infer rheological properties \citep{Ewoldt2024,Tamim2021,Edgeworth1984}. A common example is the slump test, predominantly used on construction sites to assess the workability of fresh concrete \citep{CLAYTON20033}. In this test, an open-ended mould is filled with concrete and then lifted vertically, allowing the yield stress to be inferred from the reduction in height of the resulting mound \citep{WHITE2025139839,DOMONE1998177}. Other proto-rheometric methods for measuring yield stress include flow down an inclined plane \citep{Coussot1995}, measuring the maximum bubble size that remains trapped within a fluid \citep{Hossain2024,LOPEZ2018,Pourzahedi2022}, and gravitational extension tests \citep{Geffrault2023,Geffrault}.

Both rheological and proto-rheological flows are challenging to model numerically because of the complexity of the constitutive behaviour of YSFs. While direct comparison between experiment and model is a cornerstone of research in Newtonian fluid mechanics, this methodology is rarely applied to YSFs. Microscopic and mesoscopic constitutive models, such as the soft glassy rheology model \citep{SOLLICH1997}, can capture key rheological behaviour, but they are too demanding computationally to enable numerical simulations of macroscopic flows, even for simple flow configurations. They also suffer from a large number of free parameters which cannot easily be inferred from rheological tests \citep{FIELDING2020}. By contrast, continuum constitutive models rely on a simplified description of the rheology of YSFs, but they have the advantage that they remain computationally tractable, thus making them suitable to simulate proto-rheological flows numerically. The Saramito-Herschel-Bulkley (SHB) model \citep{Saramito2009} is a widely used continuum constitutive model for YSFs where yielding is treated as a sharp, stress-based transition. Below yield, the material behaves as a linear viscoelastic solid, and above yield, plastic flow according to the Herschel-Bulkley relation is combined with viscoelastic behaviour. 
This model, along with its close relative the Saramito-Bingham model \citep{Saramito2007}, has been successively applied to elucidate a variety of flows, spanning, e.g., bubble rise \citep{Kordalis2023,MOSCHOPOULOS2021104670,Garbin2026} and bursting \citep{Balasubramanian2024}, flow past a cylinder \citep{Corrochano2026EVP,MOUSAVI2025105384}, flow through porous media \citep{Parvar2024531}, flow through contractions \citep{MOUSAVI2024105218} and drop spreading \citep{Jalaal2021,Jalaal2024}. However, these models introduce parameters that do not always have a direct physical interpretation and therefore pose challenges when attempting to match the rheological properties of model yield stress fluids used in experiments. In particular, a single value of the solvent viscosity, $\eta_s$, cannot reproduce the behaviour well above yield while also capturing the sub-yield response. In many applications, the yielding transition itself is of primary interest, making the choice of a physically realistic value of $\eta_s$ non-trivial. Alternative constitutive models have been proposed to unify the physics above and below yield without invoking a sharp transition at the yield criterion \citep{KDR2021}. While these models address some limitations of the Saramito framework, they predict unrecoverable deformation at all applied stress levels, which is inconsistent with experimental observations of yield stress fluids in parallel superposition rheometry tests, where an oscillatory stress is superimposed on a steady offset \citep{WOODBRIDGE2026}. However, given the rich behaviour of yield-stress fluids, it is important to assess constitutive models in the context of more complex flows.

In this paper, we explore the dynamics of a sessile drop subject to vertical oscillations of its substrate as a proto-rheometric equivalent of the parallel superposition test of \cite{WOODBRIDGE2026}. The drop is sufficiently large such its spreading is driven by gravitational stress in the absence of oscillatory forcing. Thus, when the substrate is oscillated vertically, the  stress applied to the droplet corresponds to the parallel superposition of steady gravitational and oscillatory components. This flow configuration provides a sensitive test of the SHB model because the vibrated drop adjusts its shape as the applied stress increases to release stress in excess of the yield stress, and thus, it remains close to the yield threshold for all values of forcing, as demonstrated by \cite{Garg2021}.

Vibrations are commonly used in the handling of YSFs to promote particle settling \citep{Pourjafar2023}, release bubbles \citep{Koch2017}, and drain tempered chocolate from moulds to create uniform coatings in chocolate manufacturing \citep{Chevalley1975}. Local vibration in the ultrasonic regime has been shown to facilitate the flow of gels, thereby reducing pumping power in processing industries \citep{PIAU2007_vibration}. At lower frequencies, \citet{Shiba2009} observed inertially driven, free-standing convection rolls in millilitre-sized drops of gels, and \citet{Wolf2015} reported the spontaneous formation of evenly spaced, persistent holes in Carbopol gels dependent on sample mass, across a range of acceleration values and frequencies. \citet{BergemannTHESIS} combined experiments and numerical simulations of a viscoplastic model to investigate the spreading of large drops of tempered chocolate subject to vertical oscillations. They observed cyclic growth and shrinkage of regions of flow and spatial distributions of the shear rate similar to those observed in non-vibrated yield-stress drops spreading under capillary forces \citep{Jalaal2021,Jalaal2024}. \citet{Garg2021} performed related experiments by applying vertical oscillations to large drops of Carbopol and tempered chocolate, with fixed size and shape, but different values of the yield stress. They reported proportionality between the dimensional threshold acceleration required to initiate spreading and the yield stress. 

In this paper, we build on the experiments of \cite{Garg2021} with a dual objective to:
\begin{enumerate}
\item Determine the extent to which current continuum models can capture complex flows. We perform a direct comparison between experiments and numerical simulations of the Cauchy equations coupled to the SHB constitutive model, using the finite volume open-source code Basilisk \citep{Popinet2015}. We demonstrate the limitations of the model and show that the choice of solvent viscosity has a significant impact on the predicted behaviour, highlighting the sensitivity of parameter selection when modelling materials near the yield point.

\item Establish the vibrated drop as a powerful proto-rheological configuration. We show that the onset of spreading can be obtained in a single drop experiment by making successive measurements for increasing values of acceleration. By quantifying the effects of drop rheology and geometry on the spreading behaviour, we deduce an empirical scaling which coins a new proto-rheometric test for measuring yield stress in systems where vibration is used to promote fluidisation.
\end{enumerate}

The outline of the paper is as follows. The experimental apparatus, material preparation and drop deposition methods are detailed in \S~\ref{sec:Experimental_methods}. The mathematical model and numerical methods are presented in \S\S~\ref{subsec: Problem description},~\ref{subsec: Initial condition and numerical method}, with a discussion of the model parametrisation choices for direct comparison with experiments in \S~\ref{subsec:model_param}. In \S~\ref{sec:Results_discussion}, we present direct comparisons between experiments and numerical simulations of sessile drops of different shapes under gravity alone (\S~\ref{subsec:sessile}), and of the dynamic response of drops to vertical oscillations of their substrate (\S\S~\ref{subsec:dynamics},~\ref{subsec:numerics_results}). Proto-rheology with oscillating drops is addressed in \S~\ref{sec:proto}, where we quantify drop spreading in terms of engineering strain measurements in \S~\ref{subsec:onset of drop spreading}, analyse the influence of drop rheology and geometry in \S~\ref{subsec:effect_rheol_geo} and demonstrate an approximate scaling of these results which yields a simple relation between yield stress and acceleration in \S~\ref{subsec:Vibrated_drop_rheometer}. Conclusions are given in \S~\ref{sec:conclusions} where we discuss the importance of assessing constitutive models in non-standard flow configurations in the light of our findings.

\section{Experimental methods}\label{sec:Experimental_methods}
\begin{figure}
    \centering
    \includegraphics[scale=0.85]{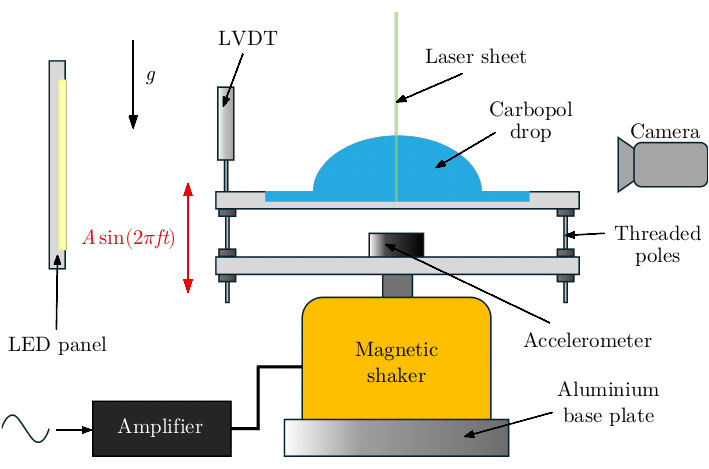}
    \caption{Schematic diagram of the vibration experiment (not to scale).}
    \label{fig:experimental_setup}
\end{figure}

The experimental system is shown schematically in figure~\ref{fig:experimental_setup}. It builds on the apparatus developed by \cite{Garg2021}, and thus we limit our description to its salient features. A drop of yield-stress fluid is initially at rest on a thin layer of the same material contained within a circular trough of diameter 60~mm and depth of $1.00\pm0.02$\,mm, which was milled into a circular Perspex plate with a diameter of 100~mm. This thin precursor layer was created prior to depositing the drop at its centre by overfilling the trough with yield-stress fluid, and removing the excess by slowly dragging a square-edged ruler across the surface at a 30\textdegree{} angle to avoid the washboard instability of the free surface \citep{Bergemann2018_wedge}. The precursor layer serves to eliminate wetting effects associated with the motion of a triple contact line, thus ensuring that the measured spreading dynamics is governed by the bulk rheology of the fluid. We discuss the choice of fluids in \S~\ref{subsec:Material_prep} and describe in \S~\ref{subsec:deposition methods} the methods used to deposit yield-stress drops of different shapes onto the substrate layer, namely cylinders, tall drops and ellipsoid caps. The substrate plate is rigidly mounted onto a second Perspex plate via three vertical threaded rods, which enable its levelling to within 0.05\textdegree{} of the horizontal using adjustable nuts. The whole assembly is fastened to the armature of a permanent-magnetic shaker (V201, LDS), oriented to impose vertical displacement. Finally, the shaker is itself bolted to a levelled heavy aluminium plate in order to suppress parasitic motion at high accelerations. 

We oscillated the substrate by driving the permanent-magnet shaker with a sinusoidal signal generated in \textsc{LabVIEW} and amplified by a linear power amplifier (LPA100, LDS). 
Experiments were performed by keeping the oscillation amplitude fixed, at values of  $A=1.2$\,mm or 2\,mm and varying the forcing frequencies in the range $8\,\mathrm{Hz}\leq f_0\leq 35\,\mathrm{Hz}$, to impose different values of the maximum acceleration, up to $a= A(2\pi f_0)^2=9.86g$, where $g$ is the acceleration due to gravity. The oscillation amplitude was calibrated as a function of the shaker input voltage at each forcing frequency using an LVDT (S015, Solartron Metrology) mounted on the substrate plate. The power of the harmonic content of the oscillating platform assembly was quantified using an accelerometer (353B43, PCB Piezotronics) mounted on the Perspex plate beneath the substrate. This was found to be less than 1\,\% of the power of the fundamental mode for frequencies $f_0\geq 8$~Hz, and thus we selected this threshold as the minimum forcing frequency. 

\begin{figure}
    \centering
    \includegraphics[scale=1.0]{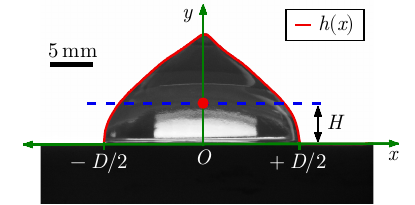}
    \caption{Image of a 3\,mL tall drop of 2.2\,g\,L$^{-1}$ Carbopol. The red line shows the outline of the drop $h(x)$ detected by \textsc{Matlab}'s Sobel function. Blue dashed line defines the height of the centre of mass. The coordinate system is shown, with the origin centre of the base of the drop.}
    \label{fig:image_processing}
\end{figure}
We monitored the drop in side-view using a CMOS high-speed camera (FLIR, Blackfly USB-3) coupled to a COSMICAR television lens ($f=75$\,mm, 1:1.4) which yields a resolution of 18.6~$\pm$~0.8~px/mm (see figure~\ref{fig:image_processing}). The camera was oriented such that the upper edge of the substrate was parallel to the edge of the image. The drop was uniformly backlit by an LED panel casting diffuse lighting. Light reflection at the surface of the drop, which reduced image quality most severely for ellipsoid-cap-shaped drops, was minimised by masking the portion of LED panel outside of the image frame. The dynamic response of the drop was captured at a rate of 286.5~frames per second. At the largest oscillation amplitude, the change in the camera viewing angle was less than 0.2\textdegree{}, such that the error on the measurement of the drop height was less than the image resolution of a pixel. For cylindrical-shaped drops the central plane of the drop could not be be accurately captured from the backlit image. Instead, we illuminated the central vertical cross section of the drop with a 1\,mm thick laser sheet from an Nd:YAG laser (Solo II-15, New Wave) coupled to a cylindrical lens, which was oriented normal to the optical axis of the camera. In all cases, the outline of the drop and substrate surface were tracked in \textsc{Matlab} using the Sobel function and the height of the centre of mass (CoM) of the drop above the substrate, $H$, was calculated from a first-moment integral of the projection of the drop onto the central $xy$ plane,
\begin{equation}
    H= \frac{1}{2} \frac{\int_{-D/2}^{+D/2} h(x)^2 \,\mathrm{d}x}{\int_{-D/2}^{+D/2} h(x) \,\mathrm{d}x},
\end{equation}
where, $D$ is the diameter of the footprint of the drop and $h(x)$ is the outline of the drop shown in red in figure~\ref{fig:image_processing}.

\subsection{Material preparation and characterisation}\label{subsec:Material_prep}

\begin{table}
  \begin{center}
  \def~{\hphantom{0}}
  \begin{tabular}{lcccccc}
      Material &  $\rho$ (kg\,m$^{-3}$)  & $\tau_0$ (Pa)  &   $k$ (Pa\,s\,$^n$) & $n$ &  $G^\prime$ (Pa) & $G^{\prime\prime}$ (Pa) \\[3pt]
       2.2\,g\,L$^{-1}$ Carbopol & $1008\,\pm25$  &  $41.4\,\pm\,0.6$ & $24.8\,\pm\,0.6$ &  $0.317\,\pm\,0.002$ & $225\,\pm\,4$ & $11.2\,\pm\,0.9$ \\
       3\,g\,L$^{-1}$ Carbopol & $1045\,\pm26$ & $51.8 \pm 1.1$ & $36.8 \pm 1.0$ & $0.278 \pm 0.011$ & $230 \pm 4$ & $16.1\,\pm\,0.7$ \\
       4\,g\,L$^{-1}$ Carbopol & $1048\,\pm26$ & $61.3 \pm 0.3$ & $35.7 \pm 0.1 $ & $0.341 \pm 0.001$ & $306 \pm 3$ & $18.7\,\pm\,0.6$ \\
       5\,g\,L$^{-1}$ Carbopol & $1049\,\pm26$ & $74.6 \pm 1.4$ & $37.5 \pm 3.4$ & $0.367 \pm 0.005$ & $343 \pm 5$ & $19.6\,\pm\,1.0$ \\
       6\,g\,L$^{-1}$ Carbopol & $1077\,\pm27$ & $94.2 \pm 0.3$ & $43.7 \pm 2.1$ & $0.358 \pm 0.013$ & $350 \pm 10$ & $19.3\,\pm\,1.1$ \\
       Body lotion & $1045\,\pm26$ & $68.9 \pm 1.1$ & $34.0 \pm 4.1$ & $0.375 \pm 0.019$ & $1260 \pm 18$ & $269\,\pm\,12$ \\
  \end{tabular}
  \caption{Physical properties of the Carbopol gels and the body lotion. The yield stress $\tau_0$, power index $n$ and the consistency index $k$ were obtained by fitting the one-dimensional Herschel–Bulkley model to flow curves. $G$ is the elastic shear modulus and $G''$ is the loss modulus obtained from the small-strain-amplitude limit of oscillatory shear rheometry.}
  \label{tab:rheology}
  \end{center}
 \end{table}

We used Carbopol gels and an emulsion (commercial body lotion Soap and Glory, Walgreens Boots Alliance, UK) as the working fluids. The Carbopol solutions were prepared at five different concentrations (listed in table~\ref{tab:rheology}) as per the protocol described in \cite{Garg2021} by adding the required mass of Carbopol Ultrez-21 powder (Lubrizol) to 0.021\,M sodium hydroxide solution prepared in deionised water. The powder was incorporated into the solution using a handheld electric mixer until no visible agglomerates remained (typically 15~min), whilst minimising the imposed shear to avoid damaging the gel network. The sample was then stirred intermittently over the following two hours and left to rest overnight to allow the Carbopol to reach a fully swollen state. The samples were degassed in a vacuum chamber and subsequently centrifuged. No additional preparation was required for the commercial body lotion. All samples were stored in sealed containers, and no measurable change in rheology was observed over the storage period. 

The rheological characterisation of the working fluids was performed using a Kinexus Pro+ rheometer (Malvern), with 20\,mm cross-hatched parallel plates to minimise wall slip, at a gap height of 1\,mm. The temperature was set to 21\textdegree{}C to match the temperature of the laboratory, $21\pm1$\textdegree{}, where the vibration experiments were performed. The Herschel-Bulkley model, $\tau = \tau_0+k\dot{\gamma}^{n}$, where $\tau$ is the stress and $\dot{\gamma}$ the shear rate, was fitted to the steady flow curves obtained from a stress-controlled ramp-down protocol to extract the yield stress $\tau_0$, power index $n$, and consistency index $k$ reported in table~\ref{tab:rheology}. The storage and loss moduli, $G^\prime$ and $G^{\prime\prime}$, were determined from the linear viscoelastic plateau in a strain amplitude sweep at $\omega=1$\,Hz, where both moduli were approximately independent of strain amplitude. The elastic modulus was taken as the storage modulus, $G=G^\prime$ within the linear viscoelastic regime. The fluid densities, $\rho$, were obtained by measuring the mass of a 10~mL sample in a syringe. All rheological measurements were performed in triplicate and the uncertainties reported in table~\ref{tab:rheology} correspond to the standard error of the measurements.
Finally, the surface tension of yield-stress fluids remains a subject of investigation. For Carbopol, values in the literature vary between approximately $\sigma = 50$\,mN\,m$^{-1}$ \citep{Geraud2014} and the surface tension of pure water. We used this lower bound value to parametrise the numerical model in this study.

\subsection{Drop deposition methods}\label{subsec:deposition methods}

\begin{figure}
    \centering
    \includegraphics[scale = 1]{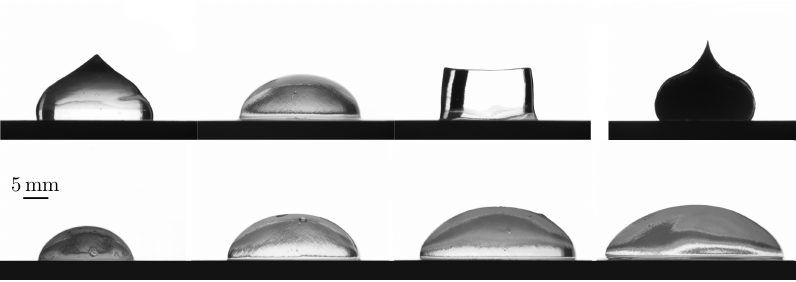}
    \caption{Drop geometries used in experiments. Top row (all 3.0\,mL), left to right: extruded drop with a pronounced tip, ellipsoid-cap-shape drop and a moulded cylindrical drop using 2.2g\,L$^{-1}$ Carbopol tall, and an extruded drop with a pronounced tip made with body lotion. Bottom row (all 2.2g\,L$^{-1}$ Carbopol ellipsoid-cap-shape drops), left to right:  with volumes 2.0\,mL, 4.0\,mL, 6.0\,mL, 8.0\,mL.}
    \label{fig:all_drops}
 \end{figure}

Inspired by the varied configurations in which yield-stress fluids are vibrated \citep{Chevalley1975,BergemannTHESIS,Schieler2001,Koch2017}, we devised three methods for depositing onto the precursor layer drops of yield-stress fluids with widely differing shapes -- moulding of a cylindrical drop, and extrusion of both a tall drop with a pronounced tip and an ellipsoid cap. Figure~\ref{fig:all_drops} shows the range of drop shapes and sizes generated from these deposition methods.

An aluminium mould (diameter $\times$ height = 20\,mm $\times$ 10\,mm) was used to produce the cylindrical drop. The mould was filled with yield-stress fluid and the surface levelled using the same procedure as for the precursor layer. The mould was then slowly lifted vertically until the drop was entirely free-standing. 
The outer diameter of the mould was 64\,mm and the underside had bevelled edges so it rested on the substrate without disturbing the precursor layer. The barrel of the mould was pierced with 10 holes to ease removal of the fluid and avoid suction of the precursor layer. 
Once the mould was removed, the free-standing cylindrical drop settled to a height of $9.9 \pm 0.2$\,mm. The volume of the deposited drop was smaller than that of the mould (3.1\,mL) as some material remained adhered to its inner surface.

Extrusion of the tall drop with a pronounced tip was achieved by a similar deposition protocol to \cite{Bergemann2018_wedge} and \cite{Garg2021}. A 10\,mL Luer-lock plastic syringe, with the nozzle enlarged to an internal diameter of $8.00\pm0.05$\,mm was filled with the desired volume of yield-stress fluid and placed in a holder with the nozzle positioned $12.0\pm 0.5\,$mm above the precursor fluid layer. The holder was mounted on three threaded rods enabling accurate levelling of the syringe holder. The fluid was extruded over approximately 3\,s resulting in an axisymmetric drop with a pronounced tip from fluid pinch-off. 

The axisymmetric ellipsoid-cap-shape drop was created by injecting fluid through a small channel in the side of the substrate connected to a vertical conduit with a diameter of 1.5\,mm exiting at the centre of the trough. The injection channel in the substrate was first prefilled with yield-stress fluid, then the precursor layer was formed and the desired volume of fluid was slowly injected through the conduit, taking care to avoid viscous buckling of the fluidised jet as it pierced through the  surface of the precursor layer. Finally, the inlet channel was sealed once the ellipsoid cap was fully formed. 

For the Carbopol gels, each of the drop geometries was prepared at each of the five concentrations listed in table~\ref{tab:rheology}, with the volume of extruded drops fixed at $V=3.0\pm 0.2\,\mathrm{mL}$. The volume of the moulded cylindrical drop was slightly smaller than the extruded drops due to the deposition method, as discussed above, giving a mean drop volume $V=2.6\pm 0.2\,\mathrm{mL}$.
For the body-lotion formulation, only tall extruded drops were considered, with $V=2.0$, $3.0$ and $4.0\,\mathrm{mL}$. In addition, for ellipsoidal caps, the volume was varied in the range $2.0\le V \le8.0\,\mathrm{mL}$ at a fixed Carbopol concentration of $2.2\,\mathrm{g\,L^{-1}}$. 
 
\section{Numerical Methods}\label{sec:numerical_methods}

\subsection{Problem description and governing equations}\label{subsec: Problem description}

\begin{figure}
    \centering
    \includegraphics[scale=1]{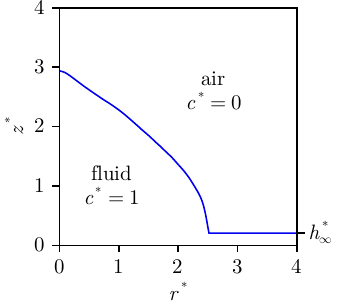}
    \caption{Schematic diagram of the computational domain used in simulations. The blue line shows the fluid-air interface, initialised as the tall extruded drop. The drop is assumed to be axisymmetric around the $z^*$ axis and the height of the precursor layer is marked $h_{\infty}^*$.}
    \label{fig:vibrated_schematic}
\end{figure} 

We consider an axisymmetric elastoviscoplastic drop on a pre-wetted surface, as shown schematically in figure~\ref{fig:vibrated_schematic}. The problem is formulated in a non-inertial reference frame attached to the substrate, such that the substrate remains stationary and its vertical oscillation is represented by a time-dependent effective acceleration consisting of gravity and a sinusoidally varying component.
For the range of drop geometries considered, we select the initial CoM height, $H_0$, as the characteristic length scale, which provides a measure of the vertical extent of the drop while accounting for variations in shape. Since the drop dynamics are driven primarily by gravity, we define the characteristic velocity, time and pressure/stress scales as ${\cal U}=\sqrt{gH_0}$, ${\cal T}=H_0/{\cal U}$ and ${\cal P}=\rho gH_0$, respectively. Denoting dimensionless variables by the superscript~$^*$, the dimensionless equations for the incompressible two-phase flow are the Cauchy equations given by
\begin{equation}\label{eq:continuity equation}
     \bnabla^* \boldsymbol{\cdot} \boldsymbol{u}^* = 0,
\end{equation}
\begin{equation} \label{eq:non dimensional navier stokes}
     \frac{\partial \boldsymbol{u}^*}{\partial t^*} + \boldsymbol{u}^* \boldsymbol{\cdot} \bnabla^* \boldsymbol{u}^* = -\bnabla^* p^* + \bnabla^* \boldsymbol{\cdot} \boldsymbol{\tau}^* + \frac{\beta}{\Rey}\bnabla^{*2}\boldsymbol{u}^* -
     \left[1+\frac{a}{g}\sin\left(2\upi \St \, t^*\right)\right]\hat{\boldsymbol{e}}_z + \frac{1}{\Bo} \kappa^* \delta_s \boldsymbol{n} ,
\end{equation}
where $\boldsymbol{u}^*$ and $p^*$ are the velocity and pressure fields, respectively, $\boldsymbol{\tau}^*$ is the extra stress tensor, 
$\hat{\boldsymbol{e}}_z$ is the unit vector in the vertical-direction, $\kappa^*$ is the curvature of the interface, $\delta_s$ is the Dirac delta function centred on the interface, and $\boldsymbol{n}$ is the outward-pointing unit normal vector to the interface. 
The term $a/g$ is the amplitude of the oscillatory acceleration component imposed on the system, relative to gravity, and $\St=f_0 \mathcal{T}$ is the dimensionless frequency of the forcing. Unlike in experiments, $a/g$ is imposed independently of the frequency of oscillations ($St$) allowing us to decouple the magnitude of imposed acceleration and the frequency.

The rheological properties of the drop (and precursor layer) are modelled with the Saramito-Herschel-Bulkley (SHB) constitutive model \citep{Saramito2009}. The dimensionless form of the SHB model obtained using the characteristic scales introduced earlier in this section is given by 
\begin{equation} \label{eq:non dimensional Saramito}
     \Wi \overset{\triangledown}{\boldsymbol{\tau^*}} + \max\!\left(0,\left( \frac{\Rey}{1 - \beta} \right) ^{1-n} \frac{|\boldsymbol{\tau}_D^*|-Y}{|\boldsymbol{\tau}_D^*|^n} \right) ^{1/n} \boldsymbol{\tau}^* = 2 \frac{1- \beta}{\Rey} {\mathsfbi{D^*}}
\end{equation}
where
\begin{equation}
    \overset{\triangledown}{\boldsymbol{\tau}^*} \equiv \frac{\mathrm{D}}{\mathrm{D}t^*}\boldsymbol{\tau}^*-\left( \bnabla^* \boldsymbol{u^*}\right)^\mathrm{T} \boldsymbol{\cdot} \boldsymbol{\tau}^*-\boldsymbol{\tau}^* \boldsymbol{\cdot} \left( \bnabla^* \boldsymbol{u}^*\right)
\end{equation}
is the upper-convected time derivative of the extra-stress tensor, and
\begin{equation}
    \mathsfbi{D}^* \equiv \frac{1}{2}\left(\bnabla^* \boldsymbol{u}^* + \left(\bnabla^* \boldsymbol{u}^* \right)^\mathrm{T} \right)
\end{equation}
is the rate-of-strain tensor. In \eqref{eq:non dimensional Saramito}, the norm $|\boldsymbol{\tau}_D^*| \equiv \sqrt{(\boldsymbol{\tau}_D^*\!\boldsymbol{:}\!\boldsymbol{\tau}_D^*)/2}$ represents the second invariant of the deviatoric component of the extra-stress tensor defined as $\boldsymbol{\tau}_D^*\equiv\boldsymbol{\tau}^* - \mathrm{tr}(\boldsymbol{\tau}^*)\mathsfbi{I}/3$, where $\mathsfbi{I}$ is the identity tensor. In addition to the forcing parameters $a/g$ and $\St$, the remaining dimensionless groups defining the system are the solvent to total viscosity ratio ($\beta$), the Reynolds number ($\Rey$), the Bond number ($\Bo$), the Weissenberg number ($\Wi$), and the yield number ($Y$), defined respectively as
\begin{equation}\label{eq:non dimensional terms}
   \beta = \frac{\eta_s}{\eta_s + \eta_p}, \;
   \Rey = \frac{\rho \mathcal{U}H_0}{\eta_s + \eta_p}, \;
   \Bo = \frac{\rho g {H_0}^2}{\mathcal{\sigma}}, \;
   \Wi = \frac{\eta_p \mathcal{U}/H_0}{G}, \;
   Y = \frac{\tau_0}{\rho g H_0},
\end{equation}
where $\eta_p = k \left(\mathcal{U}/H_0\right)^{n-1}$ denotes a characteristic polymeric viscosity based on the selected velocity and length scales, and $\eta_s$ is the solvent viscosity.
The choice of $\eta_s$, and the resulting implications for matching the sub-yield and post-yield rheology, are discussed in \S~\ref{subsec:model_param}.

\subsection{Numerical implementation and initial drop shape}
\label{subsec: Initial condition and numerical method}

Equations~\eqref{eq:continuity equation}--\eqref{eq:non dimensional Saramito} are solved using the open-source finite-volume code \textsc{Basilisk} \citep{Popinet2015}, following the axisymmetric two-phase implementation described in detail in \citet{Woodbridge2026b}. The drop--air interface is tracked using a volume-of-fluid method on an adaptive Cartesian quadtree mesh. The interface is represented by the volume-fraction field $c^*$, with $c^*=1$ in the drop phase and $c^*=0$ in the air phase, which evolves according to
\begin{equation}
    \frac{\partial c^*}{\partial t^*}
    +\bnabla^*\boldsymbol{\cdot}(c^*\boldsymbol{u}^*)=0.
\end{equation}
The momentum equation (cf. \eqref{eq:non dimensional navier stokes}) is solved in both phases using a one-fluid formulation, with the local material properties linearly interpolated using $c^*$. The density and Newtonian-viscosity ratios between the air and drop phases are both set to $10^{-3}$, for which the influence of the surrounding air on the drop dynamics is negligible.

The SHB constitutive equation (cf. \eqref{eq:non dimensional Saramito}) is integrated using the log-conformation formulation rather than by directly evolving the components of $\boldsymbol{\tau}^*$, again following the procedure adopted by \citet{Woodbridge2026b}. To adapt the existing log-conformation solver in \textsc{Basilisk} to the present elastoviscoplastic model, \eqref{eq:non dimensional Saramito} is recast in the relaxation form
\begin{equation}
    \frac{\Wi}{f+\epsilon}
    \overset{\triangledown}{\boldsymbol{\tau}^*}
    +\boldsymbol{\tau}^*
    =
    2\frac{1-\beta}{\Rey(f+\epsilon)}
    \mathsfbi{D}^*,
\end{equation}
where
\begin{equation}
    f=\max\!\left(0,\left( \frac{\Rey}{1 - \beta} \right) ^{1-n} \frac{|\boldsymbol{\tau}_D^*|-Y}{|\boldsymbol{\tau}_D^*|^n} \right) ^{1/n},
\end{equation}
and $\epsilon$ is a small numerical parameter introduced to prevent a division-by-zero error in the unyielded regime, where $f=0$. We set $\epsilon=10^{-20}$, for which the numerical results are insensitive to its value, as demonstrated in Appendix~\ref{appC}. The resulting equation is solved in terms of the logarithm of the conformation tensor, as described in detail in that study.

The initial drop interface is obtained directly from the experimentally measured profiles of the three drop shapes described in \S~\ref{sec:Experimental_methods}. For each case, the measured initial CoM height, $H_0$, taken relative to the surface of the precursor layer, is used to nondimensionalise the profile. The precursor layer has a fixed dimensional thickness $h_\infty=1$\,mm, such that $h_\infty^*=h_\infty/H_0$ varies between drop geometries; for the tall extruded drop shown in figure~\ref{fig:vibrated_schematic}, $h_\infty^*=0.20$.

The computations are performed in the square axisymmetric domain $(r^*,z^*)\in[0,4]\times[0,4]$. The mesh is initially refined to level $L=9$ and subsequently adapts over the range $5\leq L\leq9$, giving a minimum grid spacing $\Delta^*_{\min}=4/2^9=7.8\times10^{-3}$. Refinement is based on the interface curvature, and the velocity and extra-stress fields, retaining the finest resolution near the moving interface and in regions of strong velocity and stress variation while permitting coarser resolution elsewhere. Mesh-convergence tests are presented in Appendix~\ref{appC}.

Axisymmetry is imposed at $r^*=0$, with the corresponding symmetry conditions applied to the velocity, pressure, volume fraction and extra-stress fields. At the stationary substrate, $z^*=0$, no-slip and no-penetration conditions are imposed together with a zero normal gradient of the volume fraction. An open boundary condition is imposed at $z^*=4$, while free-slip and no-penetration conditions are applied at $r^*=4$. At the outer radial boundary, the volume fraction is prescribed as $c^*=1$ for $z^*\leq h_\infty^*$ and $c^*=0$ otherwise, thereby maintaining the precursor-layer thickness at $h_\infty^*$. The full componentwise specification of these boundary conditions is given by \citet{Woodbridge2026b}.

Time integration is explicit, adaptive and second-order accurate. The time step is selected adaptively using the default stability constraints in \textsc{Basilisk}~\citep{Popinet2009}, namely the Courant--Friedrichs--Lewy condition based on the face-centred velocity field, with $\mathrm{CFL}=0.8$, and the capillary-wave stability criterion associated with the explicit treatment of surface tension. A maximum time step $\Delta t^*_{\max}=10^{-4}$ is imposed, although the adaptively selected time step remains below this value in all simulations. The tolerance of the linear solver is set to $10^{-6}$.

\subsection{Model parametrisation}\label{subsec:model_param}

We considered drops of 2.2\,g/L Carbopol Ultrez 21 with $V\approx3.0$\,mL. 
We parametrised the model by calculating the dimensionless groups in \S~\ref{subsec: Problem description} based on measured values of $H_0$ and the physical properties listed in table~\ref{tab:rheology}. Steady flow curves provide accurate measurements of the Herschel-Bulkley parameters which determine the value of $\eta_p$, and the elastic modulus is measured in the linear viscoelastic regime of the unyielded fluid using oscillatory-shear measurements as discussed in \S~\ref{subsec:Material_prep}. However, $\eta_s$, and thus $\beta$, are not uniquely defined from rheometric data because of limitations of the Saramito model, which does not fully capture the rheology of elastoviscoplastic fluids such as Carbopol. Hence, the estimation of $\beta$ requires to choose whether the sub-yield or post-yield behaviour is to be captured, or alternatively, $\eta_s$ may be interpreted as the viscosity of the Newtonian solvent in which the polymer is dispersed which would lead to vanishingly small values of $\beta$.

\begin{figure}
    \centering
    \includegraphics[scale = 1]{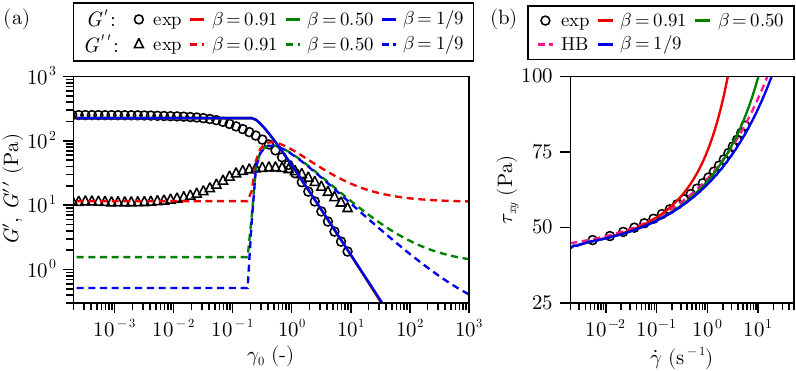}
    \caption{Rheometric characterisation of 2.2\,gL$^{-1}$ Carbopol gel for (a) oscillatory rheometry at $\omega=1$\,Hz and (b) steady-shear, both using cross-hatched parallel plate geometry. In (a) the circles show the raw data for a ramp-down flow curve and in (b) the circles show $G'$ and the triangles show $G''$. The SHB model predictions are given for $\beta=0.91$ (red), $\beta=0.50$ (green), and $\beta=1/9$ (blue), and the Herschel--Bulkley fit (magenta) is also given in (b). Note that red and green solid lines are not visible because the blue solid line superimposes exactly.}
    \label{fig:parametrisation_curves}
\end{figure}

Figure~\ref{fig:parametrisation_curves} compares our rheometric data with the one-dimensional SHB model fitted using three distinct values of the solvent viscosity, each giving rise to different values of $\beta$. The experimental data covers the range of strains and strain rates imposed in the vibrated drop experiment. The oscillatory-shear rheometry data shown in figure~\ref{fig:parametrisation_curves}(a) confirms that the value of $\beta$ does not affect the SHB prediction of the storage modulus, $G'$, as indicated by the superposition of red, green and blue solid lines. The model captures the experimental trend accurately, except near yield, where the transition in the SHB model is more abrupt than in the experiment. The abrupt yielding transition also fails to capture the smooth overshoot in $G''$ near the yield threshold, which has been a key focus of recent constitutive model development \citep{KDR2021}. This is a consequence of the Oldroyd--Prager formulation underlying the SHB model, encoded by the $\max$ function in equation \eqref{eq:non dimensional Saramito}, and is not particular to any one choice of $\eta_s$. However, regardless of discrepancies near the yield point, the SHB model data clearly indicates that $\beta$ is a key determinant for the prediction of the loss modulus, $G''$. 

For $\beta=0.91$, $\eta_s$ is obtained from the loss modulus in the linear viscoelastic regime as $\eta_s = G^{\prime\prime}/\omega=11.2$\,Pa\,s. This choice accurately captures the small-strain behaviour in the strain-amplitude sweep, and is therefore a good representation of the unyielded rheology. However, it artificially inflates the effective viscosity of the yielded material, because the model requires the loss modulus to plateau at large strains to the same value as in the linear viscoelastic regime. Hence, this parametrisation of the model fails to capture the rate of decrease of $G''$ post-yield. 

The value of $\beta=0.50$ is obtained by choosing $\eta_s=\eta_p=1.17$\,Pa\,s. In this case, the SHB model satisfactorily captures the post-yield rheology of Carbopol within the experimental range of strain values, but the sub-yield rheology is significantly misrepresented because the loss modulus is underpredicted by an order of magnitude at small strains. Finally, the value of $\beta=1/9$ is commonly adopted in the numerical literature as the lowest value of $\beta$ which enables numerical stability in finite volume methods \citep{Jalaal2024}. For Carbopol (2.2\,g/L), it corresponds to $\eta_s=0.15$\,Pa\,s. This parametrisation provides the closest agreement with our experimental data at large strain amplitudes, but severely underpredicts the loss modulus in the linear viscoelastic regime (by 1.5 orders of magnitude). 

Moreover, the inclusion of a non-zero value of $\eta_s$ also affects the prediction of the steady flow curve shown in figure~\ref{fig:parametrisation_curves}(b), which indicates yielded behaviour alone. At low shear rates, all three curves agree well with the experimental data; however, at higher shear rates the influence of solvent viscosity becomes apparent, with the divergence from the experimental data of the curve for $\beta=0.91$, which overpredicts the stress. The predictions for $\beta=0.50$ and $\beta=1/9$ provide much better agreement, yet neither agrees with the experimental data as well as the Herschel--Bulkley model. The remaining discrepancy is therefore likely associated with the non-vanishing solvent contribution in the SHB model, which remains too large to represent the yielded rheology of Carbopol under strong flows.

\section{Dynamic response of elastoviscoplastic drops to vertical vibration}\label{sec:Results_discussion}

We begin in \S~\ref{subsec:sessile} by characterising the drops after deposition, by comparison between experiments and numerical simulations of drops subject to gravity alone. All the drops considered in this study are large which means that gravitational stresses dominate over capillary stresses, as indicated by the Bond number $Bo= \rho g H_0^2/\sigma \ge 2.7$. Having established that in their sessile state, the drops comprise regions of both yielded and unyielded material, we explore the response of these two-phase drops to vertical oscillatory forcing. We characterise the dynamics of the experimental drops in \S~\ref{subsec:dynamics} and investigate the challenges of direct comparison with numerical simulations in \S~\ref{subsec:numerics_results}.

\subsection{Sessile drops under gravity}
\label{subsec:sessile}

\begin{figure}
    \centering
    \includegraphics[width=\linewidth]{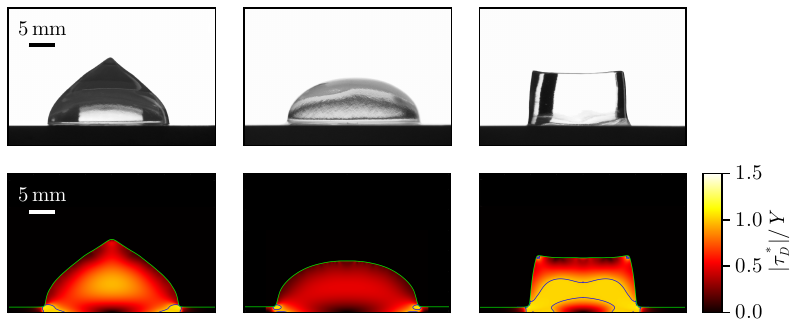}
    \caption{Sessile drop shapes. Top row: equilibrium shape after settling for 5~s of the experimental Carbopol (2.2~g/L) drops monitored in figure~\ref{fig:sessile_log}(a). Bottom row: Stress invariant maps from simulations of the corresponding drops under gravity alone ($a/g=0$) at $t=1.5$\,s for the extruded drops and $t=4$\,s for the cylindrical drop. Green solid lines denote the drop/air interface, while blue solid lines mark the the yield surfaces, $|\boldsymbol{\tau}_D^*| / Y =1$.} 
    \label{fig:sessile_exp_sim_comp}
\end{figure}
\begin{figure}
    \centering
    \includegraphics[scale=1]{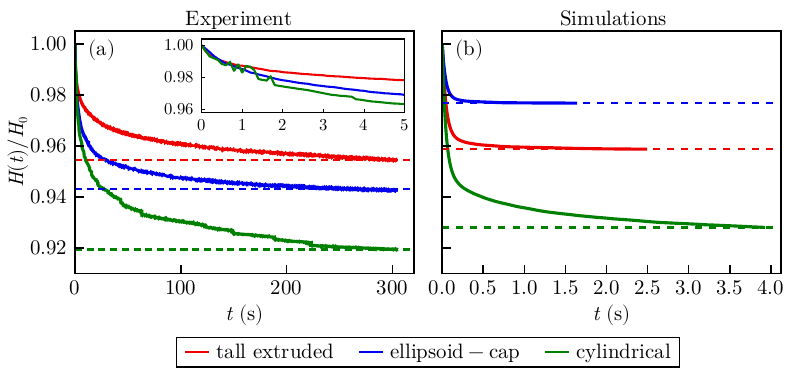}
    \caption{Settling under gravity of each of the three Carbopol (2.2~g\,L$^{-1}$) drop shapes used in experiments (left) and numerical simulations of drops subject to gravity (right), with rheology matched to that of the 2.2~g\,L$^{-1}$ Carbopol and $\beta=0.91$. Dashed lines in (a) show the height of each drop at 300\,s when the experiment ended: the tall-extruded and cap drops have nearly reached a constant height, whereas the cylindrical drop has not reached a near-constant height within the experimental time. Dashed lines in (b) show the equilibrium height of the tall-extruded and cap drops, and the height at 4\,s for the cylindrical drop. The inset in (a) shows the first 5\,s of the settling.}
    \label{fig:sessile_log}
\end{figure}

Figure~\ref{fig:sessile_exp_sim_comp} (top row) shows the three different shapes of Carbopol (2.2\,g\,L$^{-1}$) drops with $V\approx3.0$\,mL approximately 5\,s after deposition (see \S~\ref{subsec:deposition methods} for details of the deposition methods). The maximum heights of these drops are associated with hydrostatic pressure values at the drop base which all exceed the yield stress ($\tau_0=41.4\,$Pa) by at least a factor of two (133\,Pa, 84\,Pa, and 98\,Pa, for the tall-extruded, ellipsoid-cap-shaped, and cylindrical drops, respectively). Within a viscoplastic framework, commonly adopted in slump tests \citep{SAAK2004363}, we expect these values of gravitational stress to fluidise the lower portion of the drop and generate radial transport of fluid, which we refer to as spreading \citep{LIU201665}. Numerical simulations of a slumping viscoplastic cylinder by \cite{Liu2018} show that the initial yield surface rises as the drop spreads, and that the drop arrests once the yield surface coincides with the free surface of the drop. By contrast, an elastoviscoplastic drop can maintain a taller equilibrium shape because elastic stresses can balance gravitational stresses.
This is confirmed by the results of numerical simulations of the evolution under gravity of the three drop shapes, which are shown in the bottom row of figure~\ref{fig:sessile_exp_sim_comp}. Each time-simulation was initialised with the shape of the experimental drop captured 5~s after deposition and terminated once the drop height reached a plateau (i.e., the drop reached its equilibrium state). The simulations were performed with the parameters listed in table~\ref{tab:ND_parameters} which exhibit slight variations between drop shapes because of different values of $H_0$. The value of $\beta=0.91$ was chosen to match sub-yield viscoelastic properties. The bottom row of figure~\ref{fig:sessile_exp_sim_comp} shows that the distribution of the stress invariant within the central $xy$ cross-section of these drops differs significantly from the viscoplastic limit. For tall extruded drops and ellipsoid caps, the yield surface (blue line) encloses localised regions of yielded material near the outer rim where the drop merges with the precursor layer and the curvature of the free surface is large. These yielded regions are found to coexist with regions of unyielded material. They are considerably enlarged in the moulded cylindrical drop, translating into more significant shape adjustment which takes approximately 3 times longer to achieve.

Comparison between the evolution of the three drop types in the numerical simulations is shown in figure~\ref{fig:sessile_log}(b), where the relative CoM height of each drop is plotted as a function of time. Modest height reductions between approximately 2\% and 7\% over relatively short times of a few seconds point to elastic relaxation of the drop \citep{Woodbridge2026b}. We find that the extent of elastic relaxation appears to correlate with the fraction of yielded material in the drop, such that the ellipsoid cap undergoes least settling. Analogous monitoring of the evolution of the experimental drops yields very different results, as shown in figure~\ref{fig:sessile_log}(a). Although drop heights are only reduced slightly more compared to the numerical results, the settling time is more than two orders of magnitude larger suggesting a different physical mechanism. 
Because the experimental drops are formed under gravity, we do not expect significant elastic relaxation following deposition, at least in the extruded drops, and the matching of the sub-yield rheology of Carbopol should yield time scales of elastic relaxation in the moulded cylindrical drop that are similar to the numerical simulations. Thus, we hypothesise that the long-term settling of experimental drops is instead due to creep, which depends on the rheological state of the material upon deposition \citep{Lidon2017}. This is consistent with the extent of the settling of the ellipsoid cap exceeding that of the tall extruded drop, because of higher shear stresses during extrusion of the former. Creep is not modelled by the SHB constitutive relation \citep{Saramito2009,FRAGGEDAKIS2016} and thus does not feature in our numerical results. However, \citet{WOODBRIDGE2026} have shown that creep is suppressed under parallel superposition rheological tests, thus making the SHB framework appropriate to model the proto-rheological vibrated-drop configuration. 

\begin{table}
  \begin{center}
  \def~{\hphantom{0}}
  \begin{tabular}{lcccccc}
      Drop shape &  $H_0$\,(mm) & $Y$  & $Wi$  & $Re$ & $Bo$ &  $\beta$  \\[3pt]
       Tall extruded & 4.90 & 0.67 &  0.21 & 0.09 &  4.74 & 0.91 \\
       ellipsoidal-cap-shape & 3.67 &0.89 &  0.22 & 0.06 & 2.67 & 0.91 \\
       cylindrical & 4.95 & 0.66 &  0.21 & 0.09 &  4.85 & 0.91 \\
  \end{tabular}
  \caption{Non-dimensional parameters used in simulations of the three drop shapes initialised with the shape of the experimental drop captured 5\,s.}
  \label{tab:ND_parameters}
  \end{center}
 \end{table}

\subsection{Vibration experiments with a Carbopol drop}\label{subsec:dynamics}

\begin{figure}
    \centering
    \includegraphics[scale=1]{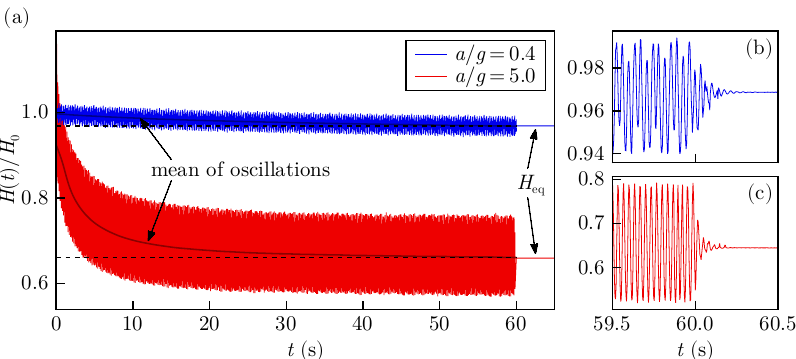}
    \caption{Drop height normalised by the initial height after deposition for a 3.0\,mL extruded drop of 2.2\,g\,L$^{-1}$ Carbopol during oscillation. (a) height during 60\,s of forcing for $a/g = 0.4$ ($A = 1.2$\,mm, $f=9$\,Hz) (blue) and $a/g = 5$ ($A = 1.2$\,mm, $f=32$\,Hz) (red). Solid dark lines: mean of the oscillations during forcing  and black dashed lines: mean height of the drop at $t=60$\,s. The  equilibrium height after forcing $H_{\mathrm{eq}}$ is labelled. Details of the viscoelastic oscillations, and subsequent decaying free oscillations once the imposed oscillations are ceased, are displayed for $a/g = 0.4$ (b) and $a/g = 5$ (c).} 
    \label{fig:Dynamics}
\end{figure}

We find that the three experimental drop types characterised in \S~\ref{subsec:sessile} exhibit qualitatively similar dynamic responses to vertical sinusoidal excitation. In figure~\ref{fig:Dynamics} we show the dynamic response of tall extruded drops of Carbopol (2.2\,g\,L$^{-1}$) for two widely different values of the forcing acceleration: $a/g = 0.4$ (blue) and $a/g=5.0$ (red) by plotting the relative CoM height of the drop as a function of time. Vertical forcing of the substrate is imposed from rest at $t=0$\,s for a duration of 60\,s, after which forcing is interrupted and the drop is left to settle under gravity.

For both values of the imposed acceleration there is an initial transient phase ($t < 20$\,s) where the mean CoM height decreases, before the rate of height reduction reaches a vanishingly small value, see figure~\ref{fig:Dynamics}(a). For $a/g=0.4$, the initial CoM height reduction is approximately linear, whereas for $a/g=5.0$ there is a rapid initial decrease over the first five seconds followed by a further gradual decrease. In both cases, the free surface of the drop deforms allowing release of stress imposed by the oscillatory forcing in excess of the yield stress, thus enabling the drop to adopt a new equilibrium shape with lower gravitational stress. The overall reduction in height is approximately 10 times larger for $a/g=5.0$ (34.1\,\%) than for $a/g = 0.4$ (3.1\,\%). The significant flattening of the drop for $a/g=5.0$, which increases its diameter by 19\,\% is associated with radial transport of fluidised material, i.e., spreading.  By contrast, at low accelerations, minor shape readjustments which increase the drop diameter by 2.5\,\% of its initial value, are consistent with alterations of the enclosed fluidised regions already present under gravity alone \citep{Woodbridge2026b}. The height reduction for $a/g=0.4$ is approximately 3\,\%, which is less than the 4.5\,\% observed for the sessile drops under gravity in \S~\ref{subsec:sessile}. This may be explained by variations in height reduction of approximately 1.5\% (at 300\,s) upon repetition of the creep experiments, which we attribute to small unavoidable differences in the extruded drop shape. However, suppression of creep under parallel superposition rheological tests demonstrated by \citet{WOODBRIDGE2026} may also contribute to the difference in height reduction of the creeping and vibrated drops.
 
The negligible height reduction for $t>20$\,s indicates that spreading is arrested and that the mean stress in the drop has either reduced below the yield stress, or at least sufficiently to make the timescale of any flow significantly longer than the period of oscillation. Hence, the drop no longer releases stress through spreading, and the work of the viscous damping forces per cycle balance the power of the forcing to yield constant-amplitude viscoelastic shape oscillations. The oscillation amplitude, which provides a measure of vertical strain at the centre of the drop, increases by a factor of approximately five with increasing acceleration, from $4\%$ of the CoM height at $a/g=0.4$ to $18\%$ at $a/g=5.0$. The CoM height following cessation of forcing ($t > 60$\,s) is approximately equal to the mean height during the oscillatory phase, indicating that the net change in height results solely from viscous spreading in our experiments. The experiments by \cite{Garg2021}, which were performed at larger amplitudes of forcing, showed some settling of the drop after interruption of forcing, indicative of elastic relaxation, and consistent with the increased asymmetry of the viscoelastic oscillations caused by the larger strains imposed on their drop.

\begin{figure}
    \centering
    \includegraphics[scale=1.0]{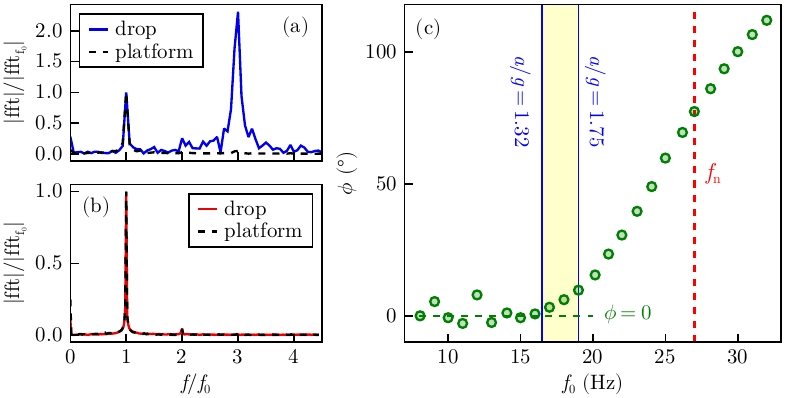}
    \caption{Fast Fourier Transform (FFT) amplitude of the drop response for (a) $f_0=9$\,Hz and (b) $f_0=32$\,Hz. The dashed black lines show the FFT amplitude of the platform signal. (c) Phase shift of the fundamental harmonic of the drop response as a function of frequency. The dashed green line shows $\phi=0$ and the red dashed line shows the natural frequency $f_\mathrm{n}=27.2 \pm 0.4$\,Hz of the drop measured from the relaxation oscillations shown in figure~\ref{fig:Dynamics}(b, c).}
    \label{fig:phase_angle_fft}
\end{figure}

The time-series extracts in figure~\ref{fig:Dynamics}(b,c) show differing periodic responses to the sinusoidal forcing of frequency $f_0$. For small frequency values ($8\,\mathrm{Hz} \le f_0 \le 15$\,Hz, figure~\ref{fig:Dynamics}(b)), the drop-height signal exhibits significant harmonic content, which gradually decays with increasing frequency and cannot be resolved for $f_0\ge 15$\,Hz (figure~\ref{fig:Dynamics}(c)). Fast Fourier Transforms (FFT) of the time series shown in figure \ref{fig:Dynamics}(b,c) indicate that for $f_0=9\,$Hz, a $3f_0$ harmonic response is excited which dominates over the $f_0$ fundamental peak (figure~\ref{fig:phase_angle_fft}(a)). By contrast, for $f_0=32$\,Hz, a single peak at the forcing frequency confirms that the drop response is harmonic (figure~\ref{fig:phase_angle_fft}(b)). 
This frequency-dependent response of the drop is observed for all the shapes investigated and is not associated with the negligible harmonic distortion of the driving signal (see \S~\ref{sec:Experimental_methods}). The amplification of the $3\! f_0$ harmonic, at $f_0= 9$~Hz, matches the natural frequency of the drop, which we measure from free oscillations following cessation of forcing as $f_\mathrm{n}=27.2 \pm 0.4$\,Hz with oscillations decaying over $0.3 \pm 0.04$\,s (figure \ref{fig:Dynamics}(b,c)). The value of $f_\mathrm{n}$ remains constant as the drop progressively flattens with increased applied acceleration, confirming it as a material property. The amplification of the $3\!f_0$ harmonic is centred around $f_0= 8$~Hz and we find a $2\!f_0$ harmonic centred around 13~Hz, although both inertial resonances exhibit broad peaks, confirming that the drop is strongly damped. 
Our results are consistent with \citet{Garg2021}'s observation of $2\!f_0$ harmonics for similar forcing frequencies. However, the larger forcing amplitudes in their experiments meant that the frequencies where superharmonic response occurred coincided with the threshold forcing acceleration required to spread the drop. The excitation in our experiment of superharmonic responses below yield is therefore suggestive of constitutive nonlinearities in the rheology of unyielded Carbopol, which are not be captured by the linear viscoelastic SHB model as we will show in \S~\ref{subsec:numerics_results}.

Figure~\ref{fig:phase_angle_fft}(c) shows the phase shift $\phi$ between the oscillations of the fundamental harmonic of the drop response extracted from the FFT and the platform oscillation as a function of frequency $f_0$ (or acceleration $a/g$). The data is collected for $t>20$\,s after the drop has finished spreading and thus oscillates viscoelastically about its new equilibrium shape. We define $\phi$ relative to the inertial phase lag of 160\textdegree{} measured for the drop response at low frequency so that $\phi\approx 0$ for the smallest value of forcing frequency. Its value remains approximately constant up to $f_0=16$\,Hz (corresponding to $a/g=1.32$). This range is where higher harmonics are excited, and can dominate the drop response, resulting in some scatter in the phase of the fundamental harmonic relative to the platform signal. For $f_0>16$\,Hz, the phase shift increases monotonically, modestly in the shaded yellow region and then more steeply, reaching 112\textdegree{} at the maximum applied forcing. This response is difficult to interpret as it couples a damped inertial resonance centred on the natural frequency $f_n$ (where $\phi$ is close to 90\textdegree{}), and the rheological response of the material to the imposed oscillatory stress relative to gravity, $a/g$, whereby an increasing phase shift of the drop response is indicative of enhanced viscous dissipation. 
Remarkably, the upper bound of the shaded yellow region also marks the forcing threshold beyond which onset of radial spreading occurs in the drop. This suggests that the growth of $\phi$ at modest frequencies is indicative of an increase in yielded material. The constant phase angle observed for $a/g < 1.32$ does not imply the absence of yielded material; rather, it indicates that the degree of yielding does not increase significantly over this range. We will further discuss the connection between phase shift of the drop response and spreading within the context of proto-rheology in \S~\ref{sec:proto}. 

\subsection{Comparison between experiments and numerical simulations of an oscillated drop}\label{subsec:numerics_results}
To gain further insight into the dynamic drop response and assess the predictive capability of the SHB model, we conducted numerical simulations of the tall extruded drop studied experimentally in \S~\ref{subsec:dynamics}, by subjecting the sessile experimental shape at $t=0$ to vertical sinusoidal acceleration superimposed on gravity. Owing to computational expense, the simulations were not run for the full 60\,s considered in the experiments; instead, results are presented here for 15\,s, with selected cases extended to 30\,s in Appendix~\ref{appB}.

\begin{figure}
    \centering
    \includegraphics[width=\linewidth]{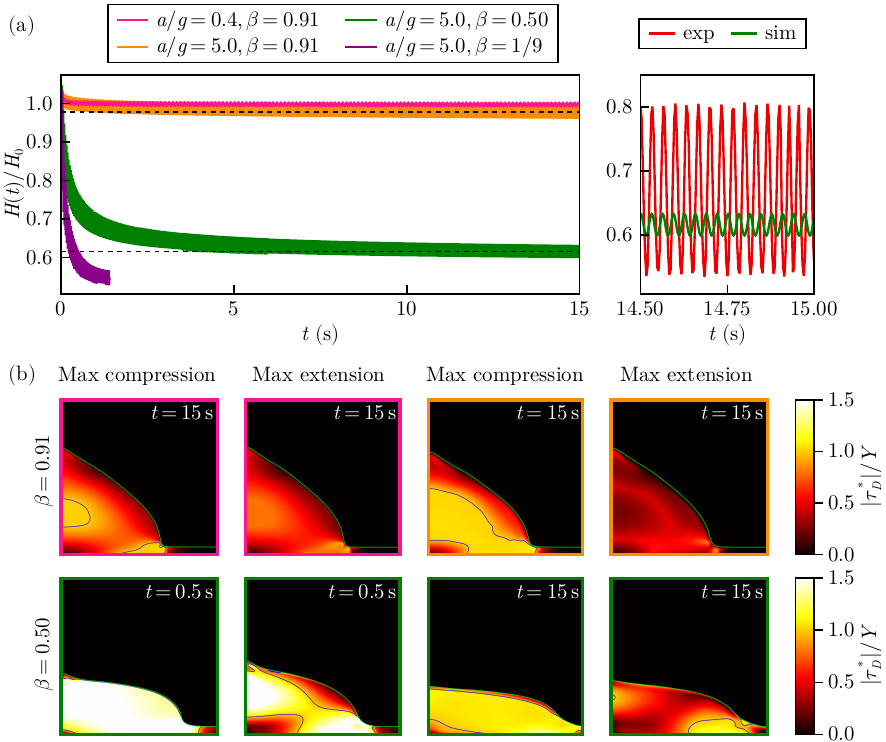}
    \caption{(a) Time series of simulations drop height for $a/g=0.4$ with $\beta = 0.91$, and $a/g=5.0$ with $\beta = 1/9,\,0.50,\,0.91$ (left) and comparison of experimental data presented in figure~\ref{fig:Dynamics} with $a/g=5.0$ with $\beta = 0.50$ (right). Black dashed lines show the mean height at $t=15$\,s. (b,c) Stress distribution for the drops shown in (a) at the maximum extension and compression of the drop in a cycle at $t=0.5$\,s and $t=15\,s$.}
    \label{fig:simulations}
\end{figure}

The magenta and orange time series of CoM height shown in figure~\ref{fig:simulations}(a) correspond, respectively, to the forcing accelerations $a/g=0.4$ and $a/g=5.0$ imposed experimentally in figure~\ref{fig:Dynamics}. In these simulations, $\beta=0.91$ as in \S\,\ref{subsec:sessile} and the value of $St$ is matched to the forcing frequencies in the experiments. 
In both cases, there is a small decrease in CoM height 
during the initial $1.5$\,s consistent with the behaviour under gravity alone shown in figure~\ref{fig:sessile_log}. This early drop-shape readjustment is not observed experimentally, because the drop is stable under gravity in the experiment. However, we find that the later time behaviour is largely unaffected by whether the drop is first allowed to settle under gravity or whether gravity and oscillatory forcing are imposed simultaneously. For consistency with the experiment, we normalise the drop height reported in figure~\ref{fig:simulations}(a) by the equilibrium height of the drop under gravity obtained from the simulations of the sessile drops subject to gravity in figure~\ref{fig:sessile_log}.

For $a/g=0.4$, the drop spreads by only $0.3\,\%$ of its initial height over $15$\,s and the amplitude of viscoelastic oscillations is only approximately $1\,\%$ of the initial height; both values are of an order of magnitude smaller than those observed experimentally. Notably, the resonance observed in the experiments at this frequency is absent in the simulations, suggesting that sub-yield rheological behaviour is not adequately captured by the SHB model. 
Furthermore, the dynamic response remains weak for $a/g=5.0$. The simulations predict only a 2\,\% decrease in the drop height over 15\,s, together with viscoelastic oscillations of amplitude approximately 4\,\% of the initial height of the drop. Hence, the extent of spreading is underpredicted by more than an order of magnitude and the oscillation amplitude is more than fivefold smaller than in the experiments. 

The stress distributions shown in figure~\ref{fig:simulations}(b) for $a/g=0.4$ (magenta) and $a/g=5.0$ (orange) are revealing. They are measured during the last cycle at $t=15$\,s when the rate of drop-height decrease has become so small that it cannot be seen on the scale of figure~\ref{fig:simulations}(a), but remains visible for $a/g=5.0$ in figure~\ref{fig:Simulation_30s}. 
For $a/g=0.4$, there are two localised fluidised regions at maximum compression. One is near the rim of the drop, driven by local capillary effects (as previously observed under gravity alone in figure~\ref{fig:sessile_exp_sim_comp}). The second is located near the centre of the drop and is caused by a combination of compressive (normal) and shear stresses generated by the motion of the free surface during viscoelastic oscillations. This second region of elevated stress, which is absent in viscoplastic simulations \citep{BergemannTHESIS,Jalaal2021}, is a feature of the elastoviscoplastic drop under compressive stress. It already appears in the sessile drop in figure~\ref{fig:sessile_exp_sim_comp}, albeit with values of stress below the yield threshold, and also in simulations of elastoviscoplastic drop spreading under gravity \citep{Jalaal2024,Woodbridge2026b}. 
At maximum extension the stress invariant falls below the yield stress throughout the drop and the material behaves as a viscoelastic solid, with a spatial stress distribution similar to that observed during compression. Thus, sustained radial spreading cannot occur because of the absence of a continuous pathway for material to flow from the drop interior to the free surface. Hence, the drop reaches a constant height about which it oscillates viscoelastically after minor reconfiguration at early times.

By contrast, the yielded regions of the drop at $a/g=5.0$ expand and merge during compression, forming a continuous channel from the centre to the rim of the drop, which enables some sustained radial spreading at early times. By $t=15$\,s, the stress invariant barely exceeds the yield threshold and thus, the spreading is essentially arrested. The weak radial spreading is due to our choice of $\beta=0.91$ which is measured sub-yield when the material behaves as a viscoelastic solid. It corresponds to $\eta_s \approx 9 \eta_p$, which artificially increases viscous dissipation in the yielded regions, thus preventing the development of flow on time scales similar to the period of oscillatory forcing. Further numerical simulations demonstrate that the extent of spreading can be increased by reducing the dimensionless frequency of forcing, thereby increasing the oscillation period, which further supports the interpretation that the larger viscosity is responsible for suppressing spreading. 

To capture the spreading observed in the experiment at large accelerations ($a/g=5.0$), we lower the value of $\beta$ and perform two further simulations  shown in figure~\ref{fig:simulations}(a): $\beta=0.50$ (green curve) and $\beta=1/9$ (purple curve). In both cases, the drop is first allowed to settle under gravity before oscillations are imposed, as the reduced solvent viscosity lowers the damping and leads to noticeable viscoelastic oscillations during settling. The free oscillations are allowed to decay before the oscillatory forcing is applied in order to avoid interference between natural oscillations and those induced by vibration. For the case $\beta=1/9$, the damping is sufficiently weak that the natural frequency can be estimated from the free oscillations, giving approximately 24\,Hz, similar to the natural frequency measured experimentally (see \S~\ref{subsec:dynamics}). 
The extent of spreading is greatly enhanced compared with the case where $\beta$ is chosen to match the sub-yield behaviour of Carbopol. With $\beta=1/9$, the mean drop height decreases to 55\,\% of its initial value within 1.5\,s as the drop spreads to the edge of the computational domain, after which further spreading is suppressed by the boundary condition imposed on $c^*$ at $r^*=4$ (see \S~\ref{subsec: Initial condition and numerical method}). We terminate the simulation at this point because the configuration differs from the experiments in which the drop radius is unbounded. 

For $\beta=0.50$, the numerical simulations approximately capture the extent of drop spreading observed experimentally. The drop spreads rapidly over the first 3 seconds and then continues to decrease gradually in height for the remainder of the simulation, reaching an approximately constant height by $t=15$\,s. This is faster than in the experiment where the drop only reaches its new equilibrium height at $t\approx 40\,$s, but the drop height $H(t)/H_0= 0.62$ determined from the mean of the oscillations near $t=15$\,s compares favourably to the experimental value of 0.66. This is because, for $\beta=0.50$, the development of flow in the yielded regions is not hindered by excessive solvent viscosity. 

The second row of images in figure~\ref{fig:simulations}(b) shows that, for $\beta=0.50$, the drop is almost entirely fluidised at the point of maximum compression during the initial rapid spreading phase ($t=0.5$\,s) and the stress invariant is significantly larger than the yield stress. The only unyielded regions that persist are located near the base of the drop, as for all simulations, and a small region at the tip of the drop, consistent with experimental observations of the persistent erect tip of tall extruded drops at all forcing accelerations. At maximum extension during the same oscillation cycle, a weakened continuous pathway for flow remains
because of the continued connection of the central and peripheral yielded regions by a narrow neck. At $t=15$\,s when the drop is essentially arrested, the stress distribution remains qualitatively similar to that observed at early spreading times; however, the magnitude of the stress invariant barely exceeds the yield threshold throughout the yielded regions such that the relatively high viscosity of the yielded material essentially suppresses further spreading as observed in the experiment.

However, the right panel of figure~\ref{fig:simulations}(a) indicates that the amplitude of the viscoelastic oscillations predicted numerically for $\beta=0.50$ remains smaller by more than five times those observed experimentally. This is a significant discrepancy and extends to all values of $\beta$ considered. This is further confirmation that close to the yield threshold, the SHB model does not accurately capture the balance between elastic storage and dissipation.

\section{Proto-rheology with oscillating drops of yield-stress fluids}\label{sec:proto} 

Despite the complex dynamics of the vertically vibrated drop discussed in \S~\ref{sec:Results_discussion}, experimental measurements suggest a threshold acceleration beyond which spreading occurs. Preliminary measurements by \cite{Garg2021} indicate that this threshold acceleration correlates with the yield stress, thus suggesting that the vibrated drop may provide a proto-rheological measure of yield stress. Having assessed the significant challenge associated with the predictive numerical modelling of the elastoviscoplastic drop response to vertical oscillatory forcing, we turn back to experiments to explore the vibrated drop as a proto-rheological tool. We show in \S~\ref{subsec:onset of drop spreading} that the threshold acceleration beyond which spreading of the drop occurs can be determined from a measure of unrecoverable strain of the drop due to oscillatory forcing. Using this metric we explore the effect of rheology and drop geometry on the onset of spreading in \S~\ref{subsec:effect_rheol_geo}. This enables us to identify an approximate scaling that relates the threshold acceleration to the yield number in \S~\ref{subsec:Vibrated_drop_rheometer}, thus providing a proto-rheometric estimate of the yield stress.

\subsection{Onset of drop spreading}\label{subsec:onset of drop spreading}

\begin{figure}
    \centering
    \includegraphics[width=0.6\linewidth]{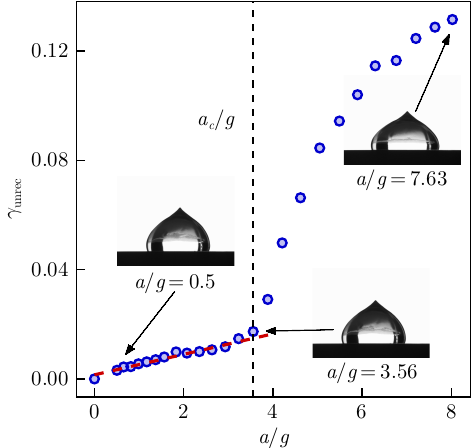}
    \caption{Unrecoverable strain as a function of imposed acceleration for a single tall extruded 5\,g\,L$^{-1}$ Carbopol drop. The imposed acceleration is increased sequentially by increasing the frequency of forcing ($f_0=8-32$) whilst keeping the amplitude constant ($A=2$\,mm). The red dashed line is a linear least-squares fitting over the first ten data points (excluding $a/g=0$) and the black dashed line indicates the threshold acceleration. Images of the drop are shown at $a/g=0.50$, $a/g=3.56$, and $a/g=7.63$.}
    \label{fig:spreading_example}
\end{figure}

To quantify the spreading behaviour of drops subject to vertical oscillation of their substrate, we performed a series of experiments on a single drop where we imposed bursts of 15\,s of oscillatory forcing followed by 5\,s rest to let the drop settle before capturing an image of its profile in the $xy$ plane. We increased $a/g$ incrementally to straddle the onset of spreading. 
Fig~\ref{fig:spreading_example} shows a typical example of such a series of experiments on a single tall extruded Carbopol drop ($5\,$g/L), where a measure of unrecoverable strain based on the total change in drop height due to oscillatory forcing,
\begin{equation}
    \gamma_\mathrm{unrec} = \frac{H_0-H_{\mathrm{eq}}}{H_0}
    \label{eq:strain1}
\end{equation}
is plotted as a function of $a/g$. Here, $H_\mathrm{eq}$ denotes the equilibrium
CoM height of the drop reached for each value successive value of $a/g$.
The data reveals two distinct settling regimes. At low forcing accelerations 
there is a small, approximately linear increase in strain with increasing acceleration, reaching $\gamma_{\mathrm{unrec}} = 0.017$ at $a/g = 3.56$.  
Stress distributions obtained from the numerical simulations (see figure~\ref{fig:simulations}) indicate that, in this regime, any unrecoverable deformation is primarily due to shape readjustment, accommodated by limited growth of yielded regions which remain localised. The inset images of the drop in figure~\ref{fig:spreading_example} show that the drop shape remains essentially unchanged prior to spreading ($a/g=0.50 \le 3.56$). \citet{Garg2021} were unable to resolve these small strain values at low accelerations and therefore concluded that no measurable height change occurred below a threshold acceleration. For $a/g > 3.56$, the unrecoverable strain increases sharply with increasing acceleration and the drop begins to spread radially. This behaviour correlates remarkably well with the increase in phase shift reported in figure~\ref{fig:phase_angle_fft}, confirming that the phase shift is indicative of enhanced viscous dissipation due to radial spreading flow, through the mechanisms elucidated by the numerical simulations in \S~\ref{subsec:numerics_results}. During the compression phase of each oscillation cycle (red curves in figures~\ref{fig:Dynamics} and green curve in figure~\ref{fig:simulations}), the drop accumulates considerable unrecoverable strain due to flow of yielded material from the centre of the drop to the free surface. At the largest accelerations shown in figure~\ref{fig:spreading_example} spreading is relatively reduced as the drop becomes increasingly flattened, thus enhancing the resistance to flow of the yielded material.

The threshold acceleration is defined as the point at which the relationship between unrecoverable strain and acceleration becomes nonlinear. A least-squares linear regression is performed over the first ten data points (excluding $a/g=0$), which reliably spans the linear regime across all experiments. The threshold acceleration is identified as the measurement point immediately preceding one with $\ge 20$\,\% relative deviation of the measured strain above the linear fit, i.e. \, $(\gamma_{\mathrm{meas}} - \gamma_{\mathrm{fit}})/\gamma_{\mathrm{fit}} \ge 0.20$; for the data shown in figure~\ref{fig:spreading_example} this gives $a_c/g = 3.56$.
For ellipsoidal-cap drops, experimental noise occasionally obscured this criterion; in such cases the threshold was identified visually as the onset of substantial strain growth, and was associated with larger uncertainty (see \S~\ref{subsec:Vibrated_drop_rheometer}). 

\subsection{Effect of rheology and geometry}\label{subsec:effect_rheol_geo}
\begin{figure}
    \centering
    \includegraphics[width=\linewidth]{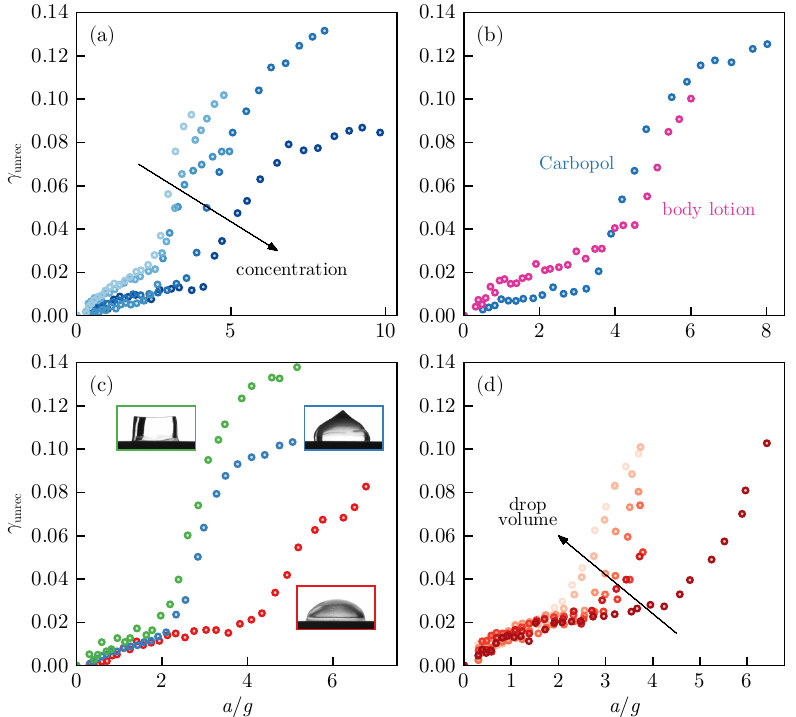}
    \caption{Unrecoverable strain as a function of imposed acceleration for (a) 3\,mL extruded Carbopol drops with concentrations of 2.2\,g\,L$^{-1}$ ($A=1.2$\,mm, $f_0=8 - 28$\,Hz), 3\,g\,L$^{-1}$ ($A=1.2$\,mm, $f_0=8 - 32$\,Hz), 4\,g\,L$^{-1}$ ($A=1.2$\,mm, $f_0=8 - 32$\,Hz), 5\,g\,L$^{-1}$ ($A=2.0$\,mm, $f_0=8 - 32$\,Hz), 6\,g\,L$^{-1}$ ($A=2.0$\,mm, $f_0=8 - 35$\,Hz); (b) 3\,mL extruded drops made with 5\,g\,L$^{-1}$ Carbopol ($A=2.0$\,mm, $f_0=8 - 32$\,Hz) and body lotion ($A=1.2$\,mm, $f_0=8 - 36$\,Hz); (c) 3\,mL drops made with 4\,g\,L$^{-1}$ Carbopol in a cylindrical shape ($A=1.2$\,mm, $f_0=8 - 32$\,Hz), extruded shape ($A=1.2$\,mm, $f_0=8 - 32$\,Hz) and a cap shape ($A=1.2$\,mm, $f_0=8 - 32$\,Hz); and (d) 2.2\,g\,L$^{-1}$ Carbopol cap-shaped drops with volumes $V = 2.0$\,mL ($A=2.0$\,mm, $f_0=8 - 37$\,Hz), $V = 3.0$\,mL ($A=1.2$\,mm, $f_0=8 - 28$\,Hz), $V = 4$.0\,mL ($A=1.2$\,mm, $f_0=8 - 28$\,Hz), $V = 5.0$\,mL ($A=1.2$\,mm, $f_0=8 - 28$\,Hz), $V = 6.0$\,mL ($A=1.2$\,mm, $f_0=8 - 28$\,Hz), $V = 7.0$\,mL ($A=1.2$\,mm, $f_0=8 - 28$\,Hz), $V = 8.0$\,mL ($A=1.2$\,mm, $f_0=8 - 28$\,Hz). Each experiment was performed in triplicate but for clarity only a single experiment is shown.}
    \label{fig:diff_plots_together}
\end{figure}
Figure~\ref{fig:diff_plots_together} shows a representative selection of the 68 experiments which were performed to quantify the spreading behaviour of a yield-stress drop under vertical vibration for different materials, drop shapes, and volumes. 
The drop response is qualitatively similar to that shown in figure~\ref{fig:spreading_example} across all experiments: a shallow, linear accumulation of unrecoverable strain at low accelerations, transitioning sharply at the threshold acceleration to steep strain increase accompanied by radial spreading. 

In figure~\ref{fig:diff_plots_together}(a), the datasets for five 3\,mL tall extruded-shape drops, one at each Carbopol concentration listed in table~\ref{tab:rheology}, are stacked in order of decreasing yield stress. The threshold acceleration increases approximately twofold across the range of yield-stress values examined, consistent with the findings of \citet{Garg2021}, who reported a linear increase in threshold acceleration with yield stress for a fixed drop shape and size. The initial strain gradient decreases with increasing Carbopol concentration, suggesting that shape readjustment at low accelerations becomes progressively weaker as the yield stress increases. This is consistent with the expected reduction in the extent of localised yielded regions at a given acceleration for an increase in the yield stress.

Figure~\ref{fig:diff_plots_together}(b) compares the spreading behaviour of 3.0\,mL extruded drops of 5\,g\,L$^{-1}$ Carbopol ($\tau_0 = 74.6$ Pa) and body lotion ($\tau_0 = 68.9$ Pa).  
Despite the different microstructural origins of the yield stress in the Carbopol (a network of cross-linked polymer chains) and body lotion (jamming of repulsive dispersed droplets) both materials exhibit similar behaviour, confirming that the  threshold acceleration is not sensitive to the microstructure \citep{Garg2021}. The body lotion has a higher threshold acceleration despite its marginally lower yield stress. We attribute this to its smaller initial height $H_0$ as shorter drops require greater acceleration to reach the same vertical stress.  
This is because the drops differ slightly in shape despite deposition using the same extrusion method and having the same volume: the body lotion exhibits a thinner, more elongated drop tip (see figure~\ref{fig:all_drops}) and a value of $H_0$ that is 4.7\% smaller than that of the Carbopol drop. The body lotion also shows a markedly larger accumulation of unrecoverable strain at the lowest accelerations ($a/g \lesssim 1$) 
suggesting significant early shape readjustment. Two geometrical and rheological factors likely contribute to this behaviour. Firstly, the higher interfacial curvature at the effective contact line of the body lotion drop (see figure~\ref{fig:all_drops}) promotes more pronounced localised yielding in this region. Secondly, \citet{Woodbridge2026b} show that higher $G$ is associated with enhanced localised yielding at accelerations below the spreading threshold, thus increasing the unrecoverable strain prior to spreading; this is the case for the body lotion ($G=1260$\,Pa) which is substantially less elastic than Carbopol, and therefore exhibits greater unrecoverable strain accumulation prior to spreading.

Figure~\ref{fig:diff_plots_together}(c) compares the spreading behaviour of 3.0\,mL ellipsoidal-cap, tall-extruded, and 2.6\,mL cylindrical Carbopol drops. There is a marked difference in the threshold accelerations where the cap-shaped drop exhibits the largest value, $a_c/g = 3.83$, whilst the extruded drop exhibits the lowest, $a_c/g = 2.16$. 
The tall-extruded and cylindrical drops share an initial centre-of-mass height $H_0 = 5.15$\,mm and exhibit comparable threshold accelerations. The ellipsoidal-cap drop, by contrast, is considerably shorter ($H_0 = 3.88$\,mm) and requires a substantially greater acceleration to initiate spreading, confirming that taller drops spread at lower threshold accelerations, consistent with the dependence on $H_0$ identified in figure~\ref{fig:diff_plots_together}(b). The post-threshold spreading of the cylindrical drop resembles a slump test: the diameter of the upper section remains approximately constant whilst the base spreads radially outwards. This behaviour has been reported previously for much larger cylindrical drops with an initial aspect ratio $h/r = 0.4$ \citep{Liu2018}. 

The influence of drop geometry on the spreading behaviour is further highlighted in Figure~\ref{fig:diff_plots_together}(d) by comparing ellipsoid caps of  2.2\,g\,L$^{-1}$ Carbopol with different volumes in the range $2.0 \le V \le 8.0 $\,mL  corresponding to CoM heights of $3.08\,\mathrm{mm} \le H_0 \le 4.52\,\mathrm{mm}$. Overall, the threshold acceleration increases as drop volume decreases. This trend is most pronounced for the 2\,mL drop which has a much smaller initial height compared with the larger volumes (an increase in height of 21\% from the 2\,mL 3\,mL to the 3\,mL drop, compared with an average of 3\% between successive larger volumes), and consequently exhibits a markedly higher threshold acceleration. For the three largest drops, the initial heights differ by less than 2\%, which leads to similar  threshold accelerations, such that the datasets overlap within experimental fluctuations over the entire range of imposed accelerations. Taken together, the data presented in figure~\ref{fig:diff_plots_together} demonstrate that both geometry and rheology govern the threshold acceleration: smaller drops with a higher yield stress require a greater acceleration to initiate spreading.

\subsection{Vibrated drop rheometer}\label{subsec:Vibrated_drop_rheometer}
\begin{figure}
    \centering
    \includegraphics[scale=1.0]{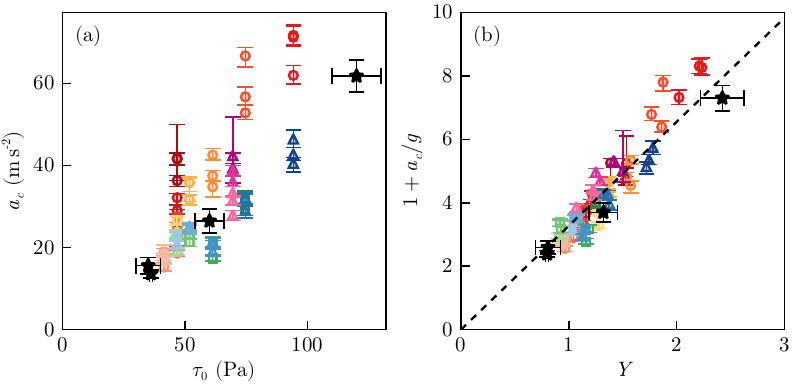}
    \caption{Threshold acceleration for all experiments as a function of (a) yield stress and (b) yield number. Each data point corresponds to a single experiment repeat. Drops made from carbopol with different geometries are distinguished by symbol shape: cap-shape (circles), cylindrical-shape (squares), and extruded-shape drops (triangles) while yield stress increases from light to dark shades. 2\,g\,L$^{-1}$ Carbopol cap-shaped drops are shown as red circles, with colour intensity increasing from dark to light with increasing volume. The emulsion is shown in pink, with darker to lighter shades corresponding to smaller to larger volumes. The black stars represent data from \citet{Garg2021}. The black dashed line in (b) is a one-parameter linear fit of the data, with a gradient of $3.3 \pm 0.25$. The vertical error bars correspond to the uncertainty in the determination of the threshold acceleration, as described in \S\,\ref{subsec:onset of drop spreading}.}
    \label{fig:ac_tau_together}
\end{figure}

The existence of a measurable threshold acceleration for the onset of drop spreading under vertical vibration offers a proto-rheometric technique for measuring the yield stress of a material that is difficult to characterise using conventional rheometers. In figure~\ref{fig:ac_tau_together}(a), we plot  the threshold accelerations $a_c/g$ obtained across all experimental parameters (see \S~\ref{subsec:effect_rheol_geo}) as a function of yield stress. We also add the results from \citet{Garg2021} for 4.5\,mL tall-extruded drops of tempered chocolate and Carbopol (2.2\,g\,L$^{-1}$, 3\,g\,L$^{-1}$ and 6\,g\,L$^{-1}$) as a benchmark. 
The threshold acceleration increases with yield stress; however, variations in drop size and shape introduce a large spread in $a_c$. To rationalise the combined influence of rheology and geometry on drop spreading, we note that the gravitational stress imposed on the drop per cycle of oscillation 
scales as $\left( a+g\right)\rho H_0$.
The use of $H_0$ is justified by the modest values of unrecoverable strain below the spreading threshold. We expect spreading to occur when the gravitational stress exceeds the yield stress, such that the threshold acceleration is given by
\begin{equation}\label{eq:scaling}
    1+a_c/g \sim \frac{\tau_0}{\rho g H_0} \equiv Y. 
\end{equation}
The yield number $Y$ encapsulates the combined influence of rheology and geometry identified experimentally in \S~\ref{subsec:effect_rheol_geo}. In particular, the inverse dependence on $H_0$ makes explicit why taller drops spread at lower accelerations despite sustaining greater gravitational loads, whilst a higher yield stress raises $Y$ and therefore $a_c$ proportionally.

Figure~\ref{fig:ac_tau_together}(b) shows that the experimental threshold acceleration approximately collapses onto a straight line when plotted as a function of $Y$. A value of $3.3\pm0.25$ is obtained for the proportionality coefficient between $ 1+a_c/g$ and $Y$ by linear fit to the data. This indicates a unique value of the ratio of yield stress to maximum imposed stress, 
$$ \frac{\tau_0}{\rho (g + a_c) H_0} = 0.30\pm0.02,$$
which is the effective Bingham number required to generate spreading. This value concurs with the theoretical predictions of \cite{LIU201665} of the critical Bingham number required for the collapse of two-dimensional viscoplastic rectangles and triangles. For rectangles with a width-to-height aspect ratio smaller than unity, they identify a critical Bingham number of 0.2646, with the critical value tending to 0.5 as the aspect ratio approaches zero. The fact that all data points have $Y \gtrsim 1$ supports the CoM height $H_0$ as the characteristic length governing the spreading behaviour, consistent with the sessile drop analysis in \S~\ref{subsec:sessile}, where the gravitational stress at the maximum drop height was shown to significantly exceed the yield stress.
This indicates that the balance between yield and gravitational stresses, with drop geometry characterised only by $H_0$ is sufficient to characterise the onset of spreading to leading order. The exact drop geometry, time-dependent forcing, microstructure, and other rheological parameters -- all of which give rise to a complex spatio-temporal stress evolution within the drop during vibration (see \S~\ref{subsec:dynamics} and \S~\ref{subsec:numerics_results}) -- play only a secondary role, and are likely responsible for the residual scatter about the linear fit in figure~\ref{fig:ac_tau_together}(b), which manifests as small variations in the numerical prefactor.

\section{Conclusion}
\label{sec:conclusions}

We have characterised the response of sessile drops of yield-stress fluid to vertical oscillations of their substrate by direct comparison between experiments and finite-volume numerical simulations of a widely used and well-regarded model of elastoviscoplastic flow, the Saramito-Herschel-Bulkley model. This model has recently been shown to capture the sub-yield behaviour of elastoviscoplastic fluids in rheological tests based on parallel superposition of steady and oscillatory components of applied stress \citep{WOODBRIDGE2026}. The proto-rheological flow configuration investigated in this paper provides a further sensitive test of the model because the vibrated drop adjusts its shape as the applied stress increases to release stress in excess of the yield stress and thus, it remains close to the the yield threshold for all values of forcing. 

Comparison between experimental and numerical sessile drops yields good agreement, indicating that all the drops settle slightly under gravity after formation. Tall extruded drops and ellipsoid caps undergo a few percent of height reduction of their centre of mass after which they reach a stable configuration, while moulded cylindrical drops slump more significantly in both experiment and numerical simulations.
We find that beyond a threshold acceleration the response of the drop to oscillatory forcing combines short-term radial flow that releases stress to return the drop to the vicinity of the yield threshold with viscoelastic shape oscillations whose amplitude depends on the imposed acceleration. The numerical simulations capture this behaviour qualitatively but also highlight limitations of the SHB constitutive model in accurately capturing key features of the dynamical response observed in the experiment.  

The numerical simulations provide insight into the spreading mechanism by giving direct access to the internal stress distribution, revealing that a continuous yielded region extending from the centre of the drop to the rim is a necessary condition for sustained spreading. When such a pathway exists, material can flow radially during the compression phase of the oscillation, allowing the drop to enlarge its footprint and release stress. We show that the rate of this flow is controlled by the parameter $\beta$, a function of the solvent viscosity $\eta_s$ which is not uniquely determined by rheological measurements. When we select $\beta=0.91$ by matching the sub-yield rheology of Carbopol in the strain amplitude sweep, we find that in the absence of oscillations, the model underpredicts both the slumping the sessile drop and its duration despite the majority of the drop being unyielded. When the drop is subject to oscillations, we find that the drop height reduction depends sensitively on the value of $\beta$. Viscous dissipation in the yielded regions is artificially increased by the large value of $\beta=0.91$ suppressing the flow of yielded material, such that the extent of spreading is better predicted by reducing $\beta$. 
The predominant choice in the literature when employing the SHB model is to assume a very small solvent viscosity \citep{Jalaal2024,Kordalis2026}; however, this does not capture the viscous response of Carbopol at small strains (see figure~\ref{fig:parametrisation_curves}), and for the vibrating drop produces entirely unphysical behaviour where the extent of spreading far exceeds that observed experimentally (see figure~\ref{fig:simulations}). In contrast with the recommendation of $\beta \le 0.1$ by \cite{Corrochano2026EVP} to best reflect the behaviour of real EVP fluid flow past a cylinder, we find that $\beta \approx 0.5$ yields closest agreement with the extent of spreading of the vibrated drop.  

Beyond the solvent viscosity, the elastic modulus $G$ presents a further source of discrepancy between simulations and experiments evidenced by the significantly smaller viscoelastic oscillation amplitudes predicted by the model, even when parametrised by matching the sub-yield regime. Our experimental observations reveal that $G$ should decrease with increasing stress, even for stresses below the yield stress, which would account for the larger viscoelastic oscillation amplitudes observed experimentally that the SHB model fails to reproduce. This is consistent with the findings of \citet{Garg2021} and \citet{WOODBRIDGE2026}, both of whom identify the need for a stress-dependent elastic modulus and present evidence that a nonlinear viscoelastic framework is appropriate for modelling Carbopol below the yield stress. Similarly, the difficulties associated with the choice of $\beta$ demonstrated here suggest that a stress-dependent solvent viscosity should also be implemented. Its value would match the sub-yield behaviour at the lowest accelerations and decrease beyond the yield stress, where the viscosity is instead governed by the characteristic polymeric viscosity. The influence of the solvent viscosity is particularly pronounced in the present study, where the ability of the material to flow depends on whether the viscous dissipation permits flow of the yielded material on the timescale of the oscillation.

The correct parametrisation of $G$ and $\eta_s$ is particularly challenging at the yield point, as the material transitions from viscoelastic solid-like to viscoelastic liquid-like behaviour. Introducing a nonlinear viscoelastic sub-yield framework could improve model parametrisation at the yield point whilst preserving the solid-like behaviour below the yield stress. This is important for complex flow scenarios such as the oscillated drop, in which yielded and unyielded regions coexist, especially when the yielded regions only marginally exceed the yield criterion. Furthermore, the higher-order harmonic response observed experimentally -- but not numerically -- for frequencies at fractions of the natural frequency of the drop 
points to a rheological origin. They offer potential for informing the viscoelastic material properties, as recently explored by \cite{Rostami2025}, although the coupling of the rheological and inertial response makes the extraction of material properties difficult. The absence of higher harmonics in the simulations suggests that it does not stem from the coexistence of yielded and unyielded regions, which are present at all applied stress levels, but rather points to limitations of the Kelvin-Voigt framework of the SHB model which constrains the steady-state response of unyielded material to be purely sinusoidal at the forcing frequency, precluding higher-order harmonics. This provides additional evidence that stress-dependent viscoelastic properties should be incorporated into the SHB model. Taken together, the discrepancies between simulations and experiments demonstrate that even well-established constitutive models such as SHB require further development -- particularly in their sub-yield description -- to capture the key features of complex EVP flows.

Whilst the full dynamic response of the drop contains rich rheological information, the experiments also identify the vibrated drop as a more powerful proto-rheometric measure of yield stress than the canonical slump test. This is because the viscoelastic response of the drop does not significantly affect the threshold acceleration at which the drop first spreads irreversibly. For the purposes of yield stress measurement, using the unrecoverable strain or the relative drop height offers a considerably simpler measurement approach than analysing the oscillatory behaviour directly. In particular, this approach only requires images of the static drop, which is practical and easily attainable, consistent with the approach of proto-rheology. We obtain the spreading threshold in a single drop experiment by making successive measurements of unrecoverable strain for increasing values of acceleration. As the vibrated-drop experiment enables access to a wide range of effective accelerations, this test is well suited to materials with high yield stresses, whilst requiring only a modest sample volume -- small enough to be practical, yet sufficiently large that $Bo\gg 1$ and spreading is gravity-driven. 

We have demonstrated that for yield-stress drops of varying size, shape, and rheology, the threshold acceleration required to induce spreading is governed primarily by the yield number $Y$ -- a balance between the yield stress and gravitational stresses -- where the centre-of-mass height $H_0$ serves as the characteristic length scale, naturally accounting for changes in drop geometry at constant volume. $H_0$ also emerges as the characteristic length scale if the drop is treated as a one-dimensional damped harmonic oscillator, making it a more appropriate length scale than the maximum height, which give rise to hydrostatic pressures in excess the yield stress, whereas the yield number for all drops using the CoM height are $Y\gtrsim 1$. Considering the complexity of the dynamic response, it is remarkable that this simple force balance provides a robust empirical scaling for the threshold acceleration for the onset of spreading in large yield-stress drops. 

Hence, the vibrated-drop rheometer can be regarded as an extension of the slump test in which the yield stress is inferred from the initial and final heights of the test sample. Results from the slump test would necessarily lie below $Y=1$ in the threshold-acceleration--yield-number plot (see figure~\ref{fig:ac_tau_together}), as they correspond to yielding driven by gravity alone, from an initial height that exceeds the static yield threshold.
The proto-rheological data presented in this paper is restricted to drops with aspect ratios $h/r \approx 1$, albeit with considerable variation in drop volume and absolute geometry. The robustness of the scaling within this regime suggests that the vibrated-drop experiment provides a reliable proto-rheological measure of yield stress for such geometries. Slender or non-axisymmetric drop geometries may produce a significantly different prefactor, and a systematic investigation of this regime warrants a future study. Incorporating additional rheological complexity into this scaling, with a view to reducing the scatter in the data, will ultimately require constitutive models that more faithfully capture the dynamics of spreading yield-stress drops -- incorporating the nonlinear sub-yield description identified in this work as necessary for accurate EVP modelling.

\begin{bmhead}[Funding] This work was funded by The University of Manchester through the Manchester Mathematical Modelling in Science and Industry initiative (AW).
\end{bmhead}

\begin{bmhead}[Acknowledgements] 
The authors gratefully acknowledge Martin Quinn for his technical support in the development of the experimental setup.
\end{bmhead}

\begin{bmhead}[Declaration of interests]
The authors report no conflict of interest.
\end{bmhead}

\begin{appen}

\section{Extended numerical simulations of oscillating drops}\label{appB}
The simulations of the oscillated drop were computationally very expensive, consequently, the main body of the paper presents results up to $t=15$\,s. For the two cases with $\beta=0.91$, the simulations were extended to $t=30$\,s to determine whether the drops continue to spread or instead reach a steady mean height about which viscoelastic oscillations persist.
\begin{figure}
    \centering
    \includegraphics[scale=1]{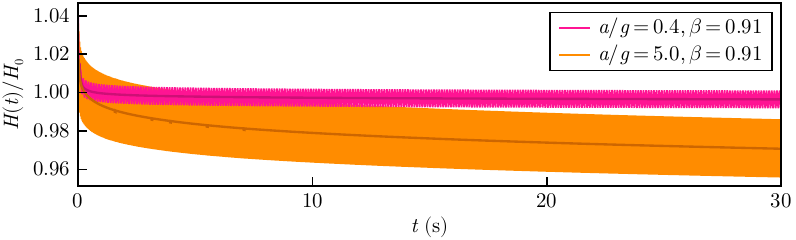}
    \caption{Time series of drop height for $a/g=0.4$ (magenta) and $a/g=5.0$ (orange), both with $\beta=0.91$. To match the frequency used to obtain those two accelerations the simulations were run with $St=0.2$ and $St=0.71$, respectively.}
    \label{fig:Simulation_30s}
\end{figure}
Figure~\ref{fig:Simulation_30s} shows that for $a/g=0.4$, the drop attains a constant mean height within approximately 5\,s and subsequently exhibits oscillations about this equilibrium. In contrast, for $a/g=5.0$, the mean height continues to decrease, albeit slowly, with an approximately linear trend over the simulated time. This difference arises because, at $a/g=5.0$, yielded regions merge during compression, enabling material to flow from the bulk to the interface and thereby producing plastic deformation. The associated timescale is long due to the high viscosity, which leads to a gradual reduction in height. Over the 30\,s simulation, the height decreases to 97.0\,\% of its initial value, compared with 97.6\,\% at 15\,s, indicating that the drop continues to evolve, albeit slowly. However, this reduction remains orders of magnitude smaller than that observed experimentally. By contrast, for $a/g = 0.4$ the change in height between 15 and 30 seconds is only 0.04\,\% confirming that the drop has effectively reached a quasi-steady state.

\section{Grid refinement and numerical regularisation dependency}\label{appC}
Here, we present convergence tests for both mesh refinement and the numerical regularisation parameter. Adaptive quadtree meshing is used in all simulations and convergence tests were performed for three levels of maximum grid refinement. In the simulations, the mesh is initially refined to the maximum level ($2^L$) throughout the domain and subsequently adapts in time to take refinement levels between $2^{L-4}$ and $2^{L}$. Refinement and coarsening are governed by eight monitored fields: the interface curvature $\kappa$, the second invariant of the deviatoric stress tensor $|\boldsymbol{\tau}_D|$, and the viscosity fields $\eta$ are assigned tolerances of $10^{-4}$, while the volume fraction $f$, velocity fields, and normal and shear components of the additional stress tensor are assigned tolerances of $10^{-3}$. We examined the effect of maximum mesh refinement on the convergence of the simulations using refinement levels $L=8$ to $L=10$ with the regularisation parameter fixed at $\epsilon = 10^{-20}$. Figure~\ref{fig:convergence}(a) shows the effect of mesh refinement on the CoM height, which shows negligible difference in the evolution of the drop height for the three levels. 
\begin{figure}
    \centering
    \includegraphics[width=\linewidth]{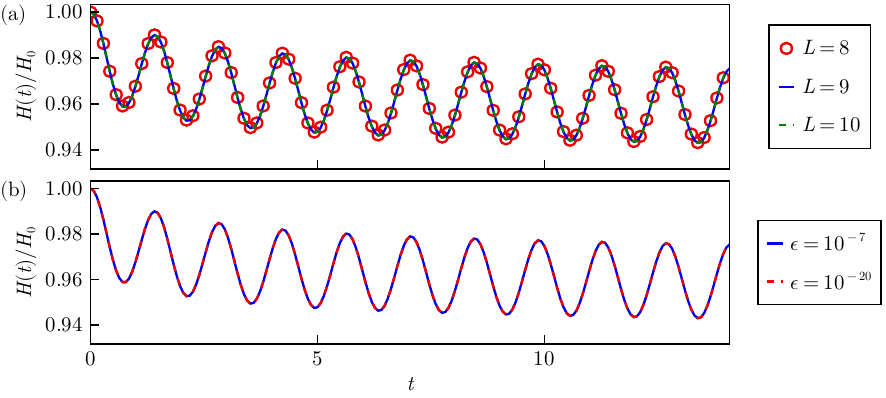}
    \caption{Convergence tests for (a) maximum mesh refinement and (b) regularisation parameter. In (a) mesh refinements of $L=8$ (red circles), $L=9$ (blue solid), $L=10$ (green dash) are compared, where the regularisation parameter is constant at $\epsilon = 10^{-20}$. In (b) the regularisation parameter is $\epsilon = 10^{-7}$ (blue solid), $\epsilon = 10^{-20}$ (red dash) are compared and the maximum mesh refinement is constant at $L=9$.}  
    \label{fig:convergence}
\end{figure}
The implementation of the SHB constitutive model in Basilisk contains a singularity below yielding conditions, which originates from the viscoplastic component of the model and renders the governing equations numerically unstable, as the viscosity diverges to infinity when $|\boldsymbol{\tau}_D^*| \leq Y$. A regularisation parameter is therefore introduced to enable the governing equations to be solved numerically. The regularisation parameter was varied between $\epsilon = 10^{-7}$ and $\epsilon = 10^{-20}$, using maximum mesh refinement level of $L=9$. The differences between the resulting solutions were indistinguishable, indicating that the simulations are insensitive to the choice of regularisation parameter over this range. All simulations presented in this study use $\epsilon = 10^{-20}$.

\end{appen}

\bibliographystyle{jfm}
 \bibliography{jfm}

\begin{thebibliography}{53}
\expandafter\ifx\csname natexlab\endcsname\relax\def\natexlab#1{#1}\fi
\def\au#1{#1} \def\ed#1{#1} \def\yr#1{#1}\def\at#1{#1}\def\jt#1{\textit{#1}} \def\bt#1{#1}\def\bvol#1{\textbf{#1}} \def\vol#1{#1} \def\pg#1{#1} \def\publ#1{#1}\def\arxiv#1{#1}\def\org#1{#1}\def\st#1{\textit{#1}}

\bibitem[Balasubramanian {\em et~al.\/}(2024)Balasubramanian, Sanjay, Jalaal, Vinuesa \& Tammisola]{Balasubramanian2024}
{\sc \au{Balasubramanian, A.~G.}, \au{Sanjay, V.}, \au{Jalaal, M.}, \au{Vinuesa, R.} \& \au{Tammisola, O.}} \yr{2024}  \at{Bursting bubble in an elastoviscoplastic medium}.  \jt{J. Fluid Mech.}  \bvol{1001},  \pg{A9}.

\bibitem[Balmforth {\em et~al.\/}(2014)Balmforth, Frigaard \& Ovarlez]{Balmforth2014}
{\sc \au{Balmforth, N.}, \au{Frigaard, I.} \& \au{Ovarlez, G.}} \yr{2014}  \at{Yielding to stress: Recent developments in viscoplastic fluid mechanics}.  \jt{Annu. Rev. Fluid Mech.}  \bvol{46},  \pg{121--146}.

\bibitem[Bergemann(2015)]{BergemannTHESIS}
{\sc \au{Bergemann, N.}} \yr{2015}  \at{Fluidisation of chocolate under vibration}. PhD thesis, University of Manchester.

\bibitem[Bergemann {\em et~al.\/}(2018)Bergemann, Juel \& Heil]{Bergemann2018_wedge}
{\sc \au{Bergemann, N.}, \au{Juel, A.} \& \au{Heil, M.}} \yr{2018}  \at{Viscous drops on a layer of the same fluid: from sinking, wedging and spreading to their long-time evolution}.  \jt{J. Fluid Mech.}  \bvol{843},  \pg{1–28}.

\bibitem[Bonn \& Denn(2009)]{BONN2009}
{\sc \au{Bonn, D.} \& \au{Denn, M.~M.}} \yr{2009}  \at{Yield stress fluids slowly yield to analysis}.  \jt{Science}  \bvol{324}~(5933),  \pg{1401--1402}.

\bibitem[Chevalley(1975)]{Chevalley1975}
{\sc \au{Chevalley, J.}} \yr{1975}  \at{Rheology of chocolate}.  \jt{J. Texture Stud.}  \bvol{6}~(2),  \pg{177--196}.

\bibitem[Clayton {\em et~al.\/}(2003)Clayton, Grice \& Boger]{CLAYTON20033}
{\sc \au{Clayton, S.}, \au{Grice, T.~G.} \& \au{Boger, D.~V.}} \yr{2003}  \at{Analysis of the slump test for on-site yield stress measurement of mineral suspensions}.  \jt{Int. J. Miner. Process.}  \bvol{70}~(1),  \pg{3--21}.

\bibitem[Corrochano {\em et~al.\/}(2026)Corrochano, Iqbal, Parvar, Le{ }{C}lainche \& Tammisola]{Corrochano2026EVP}
{\sc \au{Corrochano, A.}, \au{Iqbal, K.~T.}, \au{Parvar, S.}, \au{Le{ }{C}lainche, S.} \& \au{Tammisola, O.}} \yr{2026}  \at{The coherent structures of {EVP} fluid flow past a circular cylinder}.  \jt{Theor. Comput. Fluid Dyn.}  \bvol{40},  \pg{5}.

\bibitem[Coussot \& Boyer(1995)]{Coussot1995}
{\sc \au{Coussot, P.} \& \au{Boyer, S.}} \yr{1995}  \at{Determination of yield stress fluid behaviour from inclined plane test}.  \jt{Rheol. Acta}  \bvol{34},  \pg{534--543}.

\bibitem[Dinkgreve {\em et~al.\/}(2018)Dinkgreve, Fazilati, Denn \& Bonn]{Dinkgreve2018}
{\sc \au{Dinkgreve, M.}, \au{Fazilati, M.}, \au{Denn, M.~M.} \& \au{Bonn, D.}} \yr{2018}  \at{Carbopol: From a simple to a thixotropic yield stress fluid}.  \jt{J. Rheol.}  \bvol{62},  \pg{773--780}.

\bibitem[Domone(1998)]{DOMONE1998177}
{\sc \au{Domone, P.}} \yr{1998}  \at{The slump flow test for high-workability concrete}.  \jt{Cem. Concr. Res.}  \bvol{28}~(2),  \pg{177--182}.

\bibitem[Edgeworth {\em et~al.\/}(1984)Edgeworth, Dalton \& Parnell]{Edgeworth1984}
{\sc \au{Edgeworth, R.}, \au{Dalton, B.~J.} \& \au{Parnell, T.}} \yr{1984}  \at{The pitch drop experiment}.  \jt{Eur. J. Phys.}  \bvol{5}~(4),  \pg{198}.

\bibitem[Fielding(2020)]{FIELDING2020}
{\sc \au{Fielding, S.~M.}} \yr{2020}  \at{Elastoviscoplastic rheology and aging in a simplified soft glassy constitutive model}.  \jt{J. Rheol.}  \bvol{64}~(3),  \pg{723--738}.

\bibitem[Fraggedakis {\em et~al.\/}(2016)Fraggedakis, Dimakopoulos \& Tsamopoulos]{FRAGGEDAKIS2016}
{\sc \au{Fraggedakis, D.}, \au{Dimakopoulos, Y.} \& \au{Tsamopoulos, J.}} \yr{2016}  \at{Yielding the yield stress analysis: A thorough comparison of recently proposed elasto-visco-plastic (evp) fluid models}.  \jt{J. Non-Newton. Fluid Mech.}  \bvol{236},  \pg{104--122}.

\bibitem[Franca \& Jalaal(2024)]{Jalaal2024}
{\sc \au{Franca, H.} \& \au{Jalaal, M.}} \yr{2024}  \at{Elasto-viscoplastic spreading: From plastocapillarity to elastocapillarity}.  \jt{Phys. Rev. Res.}  \bvol{6},  \pg{013226}.

\bibitem[Garbin(2026)]{Garbin2026}
{\sc \au{Garbin, Valeria}} \yr{2026}  \at{Bubble dynamics in complex fluids}.  \jt{Phys. Rev. Fluids}  \bvol{11},  \pg{010502}.

\bibitem[Garg {\em et~al.\/}(2021)Garg, Bergemann, Smith, Heil \& Juel]{Garg2021}
{\sc \au{Garg, A.}, \au{Bergemann, N.}, \au{Smith, B.}, \au{Heil, M.} \& \au{Juel, A.}} \yr{2021}  \at{Fluidisation of yield stress fluids under vibration}.  \jt{J. Non-Newton. Fluid Mech.}  \bvol{294},  \pg{104595}.

\bibitem[Geffrault {\em et~al.\/}(2023{\natexlab{{\em a\/}}})Geffrault, Bessaies-{B}ey, Roussel \& Coussot]{Geffrault2023}
{\sc \au{Geffrault, A.}, \au{Bessaies-{B}ey, H.}, \au{Roussel, N.} \& \au{Coussot, P.}} \yr{2023{\natexlab{{\em a\/}}}}  \at{Instant yield stress measurement from falling drop size: The “syringe test”}.  \jt{J. Rheol.}  \bvol{67}~(2),  \pg{305--314}.

\bibitem[Geffrault {\em et~al.\/}(2023{\natexlab{{\em b\/}}})Geffrault, Bessaies-Bey, Roussel \& Coussot]{Geffrault}
{\sc \au{Geffrault, A.}, \au{Bessaies-Bey, H.}, \au{Roussel, N.} \& \au{Coussot, P.}} \yr{2023{\natexlab{{\em b\/}}}}  \at{Printing by yield stress fluid shaping}.  \jt{Additive Manufacturing}  \bvol{75},  \pg{103752}.

\bibitem[Géraud {\em et~al.\/}(2014)Géraud, Jørgensen, Petit, Delanoë-{A}yari, Jop \& Barentin]{Geraud2014}
{\sc \au{Géraud, B.}, \au{Jørgensen, L.}, \au{Petit, L.}, \au{Delanoë-{A}yari, H.}, \au{Jop, P.} \& \au{Barentin, C.}} \yr{2014}  \at{Capillary rise of yield-stress fluids}.  \jt{Europhys. Lett.}  \bvol{107}~(5),  \pg{58002}.

\bibitem[Hossain \& Ewoldt(2024)]{Ewoldt2024}
{\sc \au{Hossain, M.T.} \& \au{Ewoldt, R.H.}} \yr{2024}  \at{Protorheology}.  \jt{J. Rheol.}  \bvol{68},  \pg{113--144}.

\bibitem[Hossain {\em et~al.\/}(2024)Hossain, Tiwari \& Ewoldt]{Hossain2024}
{\sc \au{Hossain, T.}, \au{Tiwari, R.} \& \au{Ewoldt, R.~H.}} \yr{2024}  \at{Protorheology in practice: Avoiding misinterpretation}.  \jt{Curr. Opin. Colloid Interface Sci..}  \bvol{74},  \pg{101866}.

\bibitem[Jalaal {\em et~al.\/}(2021)Jalaal, Stoeber \& Balmforth]{Jalaal2021}
{\sc \au{Jalaal, M.}, \au{Stoeber, B.} \& \au{Balmforth, N.~J.}} \yr{2021}  \at{Spreading of viscoplastic droplets}.  \jt{J. Fluid Mech.}  \bvol{914},  \pg{A21}.

\bibitem[Kamani {\em et~al.\/}(2021)Kamani, Donley \& Rogers]{KDR2021}
{\sc \au{Kamani, K.}, \au{Donley, G.~J.} \& \au{Rogers, S.~A.}} \yr{2021}  \at{Unification of the rheological physics of yield stress fluids}.  \jt{Phys. Rev. Lett.}  \bvol{126},  \pg{218002}.

\bibitem[Koch(2017)]{Koch2017}
{\sc \au{Koch, J.}} \yr{2017}  \at{Shaking, slamming, and vibrating yield-stress fluids: Inducing particle motion in rheologically-complex materials}. PhD thesis, University of Illinois.

\bibitem[Kordalis {\em et~al.\/}(2026)Kordalis, Esposito, Zakeri, Dimakopoulos \& Tsamopoulos]{Kordalis2026}
{\sc \au{Kordalis, A.}, \au{Esposito, G.}, \au{Zakeri, P.}, \au{Dimakopoulos, Y.} \& \au{Tsamopoulos, J.}} \yr{2026}  \at{Mass transfer effects during bubble mobilisation in yield stress fluids via pressure reduction}.  \jt{J. Fluid Mech.}  \bvol{1031},  \pg{A18}.

\bibitem[Kordalis {\em et~al.\/}(2023)Kordalis, Pema, Androulakis, Dimakopoulos \& Tsamopoulos]{Kordalis2023}
{\sc \au{Kordalis, A.}, \au{Pema, D.}, \au{Androulakis, S.}, \au{Dimakopoulos, Y.} \& \au{Tsamopoulos, J.}} \yr{2023}  \at{Hydrodynamic interaction between coaxially rising bubbles in elastoviscoplastic materials: Equal bubbles}.  \jt{Phys. Rev. Fluids}  \bvol{8},  \pg{083301}.

\bibitem[Lidon {\em et~al.\/}(2017)Lidon, Villa \& Manneville]{Lidon2017}
{\sc \au{Lidon, P.}, \au{Villa, L.} \& \au{Manneville, S.}} \yr{2017}  \at{Power-law creep and residual stresses in a {Carbopol} gel}.  \jt{Rheol. Acta}  \bvol{56}~(3),  \pg{307--323}.

\bibitem[Liu {\em et~al.\/}(2016)Liu, Balmforth, Hormozi \& Hewitt]{LIU201665}
{\sc \au{Liu, Y.}, \au{Balmforth, N.J.}, \au{Hormozi, S.} \& \au{Hewitt, D.R.}} \yr{2016}  \at{Two–dimensional viscoplastic dambreaks}.  \jt{J. Non-Newton. Fluid Mech.}  \bvol{238},  \pg{65--79}, viscoplastic Fluids\: From Theory to Application 2015 (VPF6).

\bibitem[Liu {\em et~al.\/}(2018)Liu, Balmforth \& Hormozi]{Liu2018}
{\sc \au{Liu, Y.}, \au{Balmforth, N.~J.} \& \au{Hormozi, S.}} \yr{2018}  \at{Axisymmetric viscoplastic dambreaks and the slump test}.  \jt{J. Non-Newton. Fluid Mech.}  \bvol{258},  \pg{45--57}.

\bibitem[Lopez {\em et~al.\/}(2018)Lopez, Naccache \& de~Souza~Mendes]{LOPEZ2018}
{\sc \au{Lopez, W.~F.}, \au{Naccache, M.~F.} \& \au{de~Souza~Mendes, P.~R.}} \yr{2018}  \at{Rising bubbles in yield stress materials}.  \jt{J. Rheol.}  \bvol{62}~(1),  \pg{209--219}.

\bibitem[Moschopoulos {\em et~al.\/}(2021)Moschopoulos, Spyridakis, Varchanis, Dimakopoulos \& Tsamopoulos]{MOSCHOPOULOS2021104670}
{\sc \au{Moschopoulos, P.}, \au{Spyridakis, A.}, \au{Varchanis, S.}, \au{Dimakopoulos, Y.} \& \au{Tsamopoulos, J.}} \yr{2021}  \at{The concept of elasto-visco-plasticity and its application to a bubble rising in yield stress fluids}.  \jt{J. Non-Newton. Fluid Mech.}  \bvol{297},  \pg{104670}.

\bibitem[Mousavi {\em et~al.\/}(2024)Mousavi, Dimakopoulos \& Tsamopoulos]{MOUSAVI2024105218}
{\sc \au{Mousavi, M.}, \au{Dimakopoulos, Y.} \& \au{Tsamopoulos, J.}} \yr{2024}  \at{Elasto-visco-plastic flows in benchmark geometries: I. 4 to 1 planar contraction}.  \jt{J. Non-Newton. Fluid Mech.}  \bvol{327},  \pg{105218}.

\bibitem[Mousavi {\em et~al.\/}(2025)Mousavi, Dimakopoulos \& Tsamopoulos]{MOUSAVI2025105384}
{\sc \au{Mousavi, M.}, \au{Dimakopoulos, Y.} \& \au{Tsamopoulos, J.}} \yr{2025}  \at{Elasto-visco-plastic flows in benchmark geometries: Ii. flow around a confined cylinder}.  \jt{J. Non-Newton. Fluid Mech.}  \bvol{336},  \pg{105384}.

\bibitem[Ohie {\em et~al.\/}(2026)Ohie, Yoshida, Tasaka \& Murai]{Tasaka}
{\sc \au{Ohie, K.}, \au{Yoshida, T.}, \au{Tasaka, Y.} \& \au{Murai, Y.}} \yr{2026}  \at{Non-invasive inline rheometry for fluid foods containing millimeter-sized ingredients}.  \jt{J. Food. Eng.}  \bvol{404},  \pg{112744}.

\bibitem[Parvar {\em et~al.\/}(2024)Parvar, Chaparian \& Tammisola]{Parvar2024531}
{\sc \au{Parvar, S.}, \au{Chaparian, E.} \& \au{Tammisola, O.}} \yr{2024}  \at{General hydrodynamic features of elastoviscoplastic fluid flows through randomised porous media}.  \jt{Theor. Comput. Fluid Dyn.}  \bvol{38}~(4),  \pg{531 – 544}.

\bibitem[Piau \& Piau(2007)]{PIAU2007_vibration}
{\sc \au{Piau, M.} \& \au{Piau, J.M.}} \yr{2007}  \at{Wall vibrations and yield stress–shear thinning coupling (small vibrational inertia)}.  \jt{J. Non-Newton. Fluid Mech.}  \bvol{144}~(2),  \pg{59--72}.

\bibitem[Popinet(2009)]{Popinet2009}
{\sc \au{Popinet, S.}} \yr{2009}  \at{An accurate adaptive solver for surface-tension-driven interfacial flows}.  \jt{J. Comput. Phys.}  \bvol{228}~(16),  \pg{5838--5866}.

\bibitem[Popinet(2015)]{Popinet2015}
{\sc \au{Popinet, S.}} \yr{2015}  \at{A quadtree-adaptive multigrid solver for the serre–green–naghdi equations}.  \jt{J. Comput. Phys.}  \bvol{302},  \pg{336--358}.

\bibitem[Pourjafar-Chelikdani {\em et~al.\/}(2023)Pourjafar-Chelikdani, Taghilou, Rezaee, Khabazi, Taghavi \& Sadeghy]{Pourjafar2023}
{\sc \au{Pourjafar-Chelikdani, M.}, \au{Taghilou, B.}, \au{Rezaee, T.}, \au{Khabazi, N.~P.}, \au{Taghavi, S.~M.} \& \au{Sadeghy, K.}} \yr{2023}  \at{Settling dynamics of circular particles in vibrating tanks filled with a yield-stress liquid}.  \jt{Phys. Fluids}  \bvol{35},  \pg{053328}.

\bibitem[Pourzahedi {\em et~al.\/}(2022)Pourzahedi, Chaparian, Roustaei \& Frigaard]{Pourzahedi2022}
{\sc \au{Pourzahedi, A.}, \au{Chaparian, E.}, \au{Roustaei, A.} \& \au{Frigaard, I.~A.}} \yr{2022}  \at{Flow onset for a single bubble in a yield-stress fluid}.  \jt{J. Fluid Mech.}  \bvol{933},  \pg{A21}.

\bibitem[Rostami {\em et~al.\/}(2025)Rostami, Castrej\'{o}n-Pita \& Auernhammer]{Rostami2025}
{\sc \au{Rostami, P.}, \au{Castrej\'{o}n-Pita, A.} \& \au{Auernhammer, G.}} \yr{2025} Rheological insights from the oscillation dynamics of viscoelastic sessile drops,  \arxiv{arXiv: 2509.12006}.

\bibitem[Saak {\em et~al.\/}(2004)Saak, Jennings \& Shah]{SAAK2004363}
{\sc \au{Saak, A.~W.}, \au{Jennings, H.~M.} \& \au{Shah, S.~P.}} \yr{2004}  \at{A generalized approach for the determination of yield stress by slump and slump flow}.  \jt{Cem. Concr. Res.}  \bvol{34}~(3),  \pg{363--371}.

\bibitem[Saramito(2007)]{Saramito2007}
{\sc \au{Saramito, P.}} \yr{2007}  \at{A new constitutive equation for elastoviscoplastic fluid flows}.  \jt{J. Non-Newton. Fluid Mech.}  \bvol{145}~(1),  \pg{1--14}.

\bibitem[Saramito(2009)]{Saramito2009}
{\sc \au{Saramito, P.}} \yr{2009}  \at{A new elastoviscoplastic model based on the {H}erschel–{B}ulkley viscoplastic model}.  \jt{J. Non-Newton. Fluid Mech.}  \bvol{158}~(1),  \pg{154--161}.

\bibitem[Schleier-Smith \& Stone(2001)]{Schieler2001}
{\sc \au{Schleier-Smith, J.~M.} \& \au{Stone, H.~A.}} \yr{2001}  \at{Convection, heaping, and cracking in vertically vibrated granular slurries}.  \jt{Phys. Rev. Lett.}  \bvol{86},  \pg{3016--3019}.

\bibitem[Shiba {\em et~al.\/}(2007)Shiba, Ruppert-Felsot, Takahashi, Murayama, Ouyang \& Sano]{Shiba2009}
{\sc \au{Shiba, H.}, \au{Ruppert-Felsot, J.~E.}, \au{Takahashi, Y.}, \au{Murayama, Y.}, \au{Ouyang, Q.} \& \au{Sano, M.}} \yr{2007}  \at{Elastic convection in vibrated viscoplastic fluids}.  \jt{Phys. Rev. Lett.}  \bvol{98},  \pg{044501}.

\bibitem[Sollich {\em et~al.\/}(1997)Sollich, Lequeux, H\'ebraud \& Cates]{SOLLICH1997}
{\sc \au{Sollich, P.}, \au{Lequeux, F.}, \au{H\'ebraud, P.} \& \au{Cates, M.~E.}} \yr{1997}  \at{Rheology of soft glassy materials}.  \jt{Phys. Rev. Lett.}  \bvol{78},  \pg{2020--2023}.

\bibitem[Tamim \& Bostwick(2021)]{Tamim2021}
{\sc \au{Tamim, S.~I.} \& \au{Bostwick, J.~B.}} \yr{2021}  \at{Oscillations of a soft viscoelastic drop}.  \jt{npj Microgravity}  \bvol{7}.

\bibitem[White \& Lees(2025)]{WHITE2025139839}
{\sc \au{White, C.} \& \au{Lees, J.~M.}} \yr{2025}  \at{The concrete slump test—a rheometer by definition?}  \jt{Constr. Build. Mater.}  \bvol{462},  \pg{139839}.

\bibitem[Wolf {\em et~al.\/}(2015)Wolf, Dungan, McCarthy, Lim \& Phillips]{Wolf2015}
{\sc \au{Wolf, J.~M.}, \au{Dungan, S.~R.}, \au{McCarthy, M.~J.}, \au{Lim, V.} \& \au{Phillips, R.~J.}} \yr{2015}  \at{Vibration-induced geometric patterns of persistent holes in carbopol gels}.  \jt{J. Non-Newton. Fluid Mech.}  \bvol{220},  \pg{99--107}.

\bibitem[Woodbridge {\em et~al.\/}(2026{\natexlab{{\em a\/}}})Woodbridge, Amini, Lundell, Tammisola, Juel, Poole \& Fonte]{WOODBRIDGE2026}
{\sc \au{Woodbridge, A.}, \au{Amini, K.}, \au{Lundell, F.}, \au{Tammisola, O.}, \au{Juel, A.}, \au{Poole, R.~J.} \& \au{Fonte, C.~P.}} \yr{2026{\natexlab{{\em a\/}}}}  \at{Subyield dynamics in yield-stress materials}.  \jt{Phys. Rev. Lett.}  \bvol{136},  \pg{164001}.

\bibitem[Woodbridge {\em et~al.\/}(2026{\natexlab{{\em b\/}}})Woodbridge, O'Rourke, Fonte \& Juel]{Woodbridge2026b}
{\sc \au{Woodbridge, A.}, \au{O'Rourke, P.}, \au{Fonte, C.~P.} \& \au{Juel, A.}} \yr{2026{\natexlab{{\em b\/}}}}  \at{Yielded-region connectivity governs the onset of gravity-driven spreading in elastoviscoplastic drops}.  \jt{arXiv preprint arXiv:2609.01907} .

\end{thebibliography}
 
\end{document}